\documentclass[10pt,aps,prb,twocolumn,superscriptaddress,showpacs,preprintnumbers,citeautoscript]{revtex4-1}
\usepackage{times}
\usepackage{amsfonts,amssymb,amsmath}
\usepackage{color}
\usepackage{graphicx}
\usepackage{dcolumn}
\usepackage{bm}
\usepackage[colorlinks,bookmarks=false,citecolor=blue,linkcolor=red,urlcolor=blue]{hyperref}
\usepackage{ifthen}
\newlength{\abstractdiagwidth}
\newcommand{\diag}[2][]{\parbox[c]{#2\abstractdiagwidth}{\includegraphics*[width=#2\abstractdiagwidth]{#1}}}%
\newlength{\specdiagwidth}
\newcommand{\specdiag}[2][]{\parbox[c]{#2\specdiagwidth}{\includegraphics*[width=#2\specdiagwidth]{#1}}}%
\def\hds{\parbox[c]{0.8mm}{\rule{0.8mm}{0pt}}}%
\newcommand{\gd}[1]%
{%
\ifthenelse{\equal{#1}{1}}{(\diag[./GenericDiagrams/Diag_1_1]{10})}{}%
\ifthenelse{\equal{#1}{1f}}{(\diag[./GenericDiagrams/Diag_1_1_fused]{10})}{}%
\ifthenelse{\equal{#1}{2}}{(\diag[./GenericDiagrams/Diag_2_2]{27})}{}%
\ifthenelse{\equal{#1}{11}}{(\diag[./GenericDiagrams/Diag_2_1-1]{27})}{}%
\ifthenelse{\equal{#1}{3a}}{(\diag[./GenericDiagrams/Diag_3_3a]{44})}{}%
\ifthenelse{\equal{#1}{3b}}{(\,\diag[./GenericDiagrams/Diag_3_3b]{27}\,)}{}%
\ifthenelse{\equal{#1}{21}}{(\diag[./GenericDiagrams/Diag_3_2-1]{44})}{}%
\ifthenelse{\equal{#1}{111}}{(\diag[./GenericDiagrams/Diag_3_1-1-1]{44})}{}%
\ifthenelse{\equal{#1}{4}}{(\diag[./GenericDiagrams/Diag_4_4]{61})}{}%
\ifthenelse{\equal{#1}{211}}{(\diag[./GenericDiagrams/Diag_4_2-1-1]{61})}{}%
\ifthenelse{\equal{#1}{22}}{(\diag[./GenericDiagrams/Diag_4_2-2]{61})}{}%
\ifthenelse{\equal{#1}{31}}{(\diag[./GenericDiagrams/Diag_4_3-1]{61})}{}%
\ifthenelse{\equal{#1}{1111}}{(\diag[./GenericDiagrams/Diag_4_1-1-1-1]{61})}{}%
\ifthenelse{\equal{#1}{NConnected}}{(\diag[./GenericDiagrams/Diag_n_connected]{51})}{}%
\ifthenelse{\equal{#1}{NFullyConnected}}{\left(\diag[./GenericDiagrams/Diag_n_fully_connected]{43.552}\right)}{}%
\ifthenelse{\equal{#1}{NDisconnected}}{(\diag[./GenericDiagrams/Diag_n_disconnected]{51})}{}%
\ifthenelse{\equal{#1}{NDisconnectedFused}}{(\diag[./GenericDiagrams/Diag_n_disconnected_fused]{51})}{}%
\ifthenelse{\equal{#1}{DotCircle}}{(\diag[./GenericDiagrams/Diag_dotcircle]{20.2})}{}%
\ifthenelse{\equal{#1}{CircleDot}}{(\diag[./GenericDiagrams/Diag_circledot]{20.2})}{}%
\ifthenelse{\equal{#1}{DotCircleDot}}{(\diag[./GenericDiagrams/Diag_dotcircledot]{30.4})}{}%
\ifthenelse{\equal{#1}{Dot}}{(\hds\diag[./GenericDiagrams/Diag_dot]{6.4}\hds)}{}%
}%
\newcommand{\bgd}[1]%
{%
\ifthenelse{\equal{#1}{1}}{\diag[./GenericDiagrams/Diag_1_1]{10}}{}%
\ifthenelse{\equal{#1}{1f}}{\diag[./GenericDiagrams/Diag_1_1_fused]{10}}{}%
\ifthenelse{\equal{#1}{2}}{\diag[./GenericDiagrams/Diag_2_2]{27}}{}%
\ifthenelse{\equal{#1}{11}}{\diag[./GenericDiagrams/Diag_2_1-1]{27}}{}%
\ifthenelse{\equal{#1}{3a}}{\diag[./GenericDiagrams/Diag_3_3a]{44}}{}%
\ifthenelse{\equal{#1}{3b}}{\diag[./GenericDiagrams/Diag_3_3b]{27}}{}%
\ifthenelse{\equal{#1}{21}}{\diag[./GenericDiagrams/Diag_3_2-1]{44}}{}%
\ifthenelse{\equal{#1}{111}}{\diag[./GenericDiagrams/Diag_3_1-1-1]{44}}{}%
\ifthenelse{\equal{#1}{4}}{\diag[./GenericDiagrams/Diag_4_4]{61}}{}%
\ifthenelse{\equal{#1}{211}}{\diag[./GenericDiagrams/Diag_4_2-1-1]{61}}{}%
\ifthenelse{\equal{#1}{22}}{\diag[./GenericDiagrams/Diag_4_2-2]{61}}{}%
\ifthenelse{\equal{#1}{31}}{\diag[./GenericDiagrams/Diag_4_3-1]{61}}{}%
\ifthenelse{\equal{#1}{1111}}{\diag[./GenericDiagrams/Diag_4_1-1-1-1]{61}}{}%
\ifthenelse{\equal{#1}{NConnected}}{\diag[./GenericDiagrams/Diag_n_connected]{51}}{}%
\ifthenelse{\equal{#1}{NFullyConnected}}{\diag[./GenericDiagrams/Diag_n_fully_connected]{43.552}}{}%
\ifthenelse{\equal{#1}{NDisconnected}}{\diag[./GenericDiagrams/Diag_n_disconnected]{51}}{}%
\ifthenelse{\equal{#1}{NDisconnectedFused}}{\diag[./GenericDiagrams/Diag_n_disconnected_fused]{51}}{}%
\ifthenelse{\equal{#1}{DotCircle}}{\diag[./GenericDiagrams/Diag_dotcircle]{20.2}}{}%
\ifthenelse{\equal{#1}{CircleDot}}{\diag[./GenericDiagrams/Diag_circledot]{20.2}}{}%
\ifthenelse{\equal{#1}{DotCircleDot}}{\diag[./GenericDiagrams/Diag_dotcircledot]{30.4}}{}%
\ifthenelse{\equal{#1}{Dot}}{\hds\diag[./GenericDiagrams/Diag_dot]{6.4}\hds}{}%
\ifthenelse{\equal{#1}{ElemKin}}{\diag[./GenericDiagrams/Diag_elem_kin]{17}}{}%
\ifthenelse{\equal{#1}{ElemPot}}{\diag[./GenericDiagrams/Diag_elem_pot]{17}}{}%
\ifthenelse{\equal{#1}{ElemKinKin}}{\diag[./GenericDiagrams/Diag_elem_kin_kin]{34}}{}%
\ifthenelse{\equal{#1}{ElemPotPot}}{\diag[./GenericDiagrams/Diag_elem_pot_pot]{34}}{}%
\ifthenelse{\equal{#1}{ElemKinPot}}{\diag[./GenericDiagrams/Diag_elem_kin_pot]{34}}{}%
}%
\newcommand{\kag}[1]%
{%
\ifthenelse{\equal{#1}{d_1}}{\specdiag[./KagomeLattice/Diag_d_1]{110}}{}%
\ifthenelse{\equal{#1}{d_2}}{\specdiag[./KagomeLattice/Diag_d_2]{156}}{}%
\ifthenelse{\equal{#1}{hex1}}{\specdiag[./KagomeLattice/Diag_hex1]{154}}{}%
\ifthenelse{\equal{#1}{hex2}}{\specdiag[./KagomeLattice/Diag_hex2]{154}}{}%
\ifthenelse{\equal{#1}{2_1}}{\specdiag[./KagomeLattice/Diag_2_1]{154}}{}%
\ifthenelse{\equal{#1}{2_2}}{\specdiag[./KagomeLattice/Diag_2_2]{154}}{}%
\ifthenelse{\equal{#1}{3_1}}{\specdiag[./KagomeLattice/Diag_3_1]{204}}{}%
\ifthenelse{\equal{#1}{3_2}}{\specdiag[./KagomeLattice/Diag_3_2]{179}}{}%
\ifthenelse{\equal{#1}{3_3}}{\specdiag[./KagomeLattice/Diag_3_3]{154}}{}%
\ifthenelse{\equal{#1}{3_4}}{\specdiag[./KagomeLattice/Diag_3_4]{204}}{}%
\ifthenelse{\equal{#1}{3_5}}{\specdiag[./KagomeLattice/Diag_3_5]{179}}{}%
\ifthenelse{\equal{#1}{3_6}}{\specdiag[./KagomeLattice/Diag_3_6]{154}}{}%
\ifthenelse{\equal{#1}{4_1}}{\specdiag[./KagomeLattice/Diag_4_1]{254}}{}%
\ifthenelse{\equal{#1}{4_2}}{\specdiag[./KagomeLattice/Diag_4_2]{204}}{}%
\ifthenelse{\equal{#1}{4_3}}{\specdiag[./KagomeLattice/Diag_4_3]{216.5}}{}%
\ifthenelse{\equal{#1}{4_4}}{\specdiag[./KagomeLattice/Diag_4_4]{191.5}}{}%
\ifthenelse{\equal{#1}{4_5}}{\specdiag[./KagomeLattice/Diag_4_5]{254}}{}%
\ifthenelse{\equal{#1}{4_6}}{\specdiag[./KagomeLattice/Diag_4_6]{204}}{}%
\ifthenelse{\equal{#1}{4_7}}{\specdiag[./KagomeLattice/Diag_4_7]{216.5}}{}%
\ifthenelse{\equal{#1}{4_8}}{\specdiag[./KagomeLattice/Diag_4_8]{191.5}}{}%
\ifthenelse{\equal{#1}{4_9}}{\specdiag[./KagomeLattice/Diag_4_9]{254}}{}%
\ifthenelse{\equal{#1}{5_1}}{\specdiag[./KagomeLattice/Diag_5_1]{254}}{}%
\ifthenelse{\equal{#1}{5_2}}{\specdiag[./KagomeLattice/Diag_5_2]{279}}{}%
\ifthenelse{\equal{#1}{5_3}}{\specdiag[./KagomeLattice/Diag_5_3]{254}}{}%
\ifthenelse{\equal{#1}{5_4}}{\specdiag[./KagomeLattice/Diag_5_4]{266.5}}{}%
\ifthenelse{\equal{#1}{5_5}}{\specdiag[./KagomeLattice/Diag_5_5]{229}}{}%
\ifthenelse{\equal{#1}{5_6}}{\specdiag[./KagomeLattice/Diag_5_6]{191.5}}{}%
\ifthenelse{\equal{#1}{5_7}}{\specdiag[./KagomeLattice/Diag_5_7]{179}}{}%
\ifthenelse{\equal{#1}{5_8}}{\specdiag[./KagomeLattice/Diag_5_8]{204}}{}%
\ifthenelse{\equal{#1}{5_9}}{\specdiag[./KagomeLattice/Diag_5_9]{254}}{}%
\ifthenelse{\equal{#1}{5_10}}{\specdiag[./KagomeLattice/Diag_5_10]{279}}{}%
\ifthenelse{\equal{#1}{5_11}}{\specdiag[./KagomeLattice/Diag_5_11]{254}}{}%
\ifthenelse{\equal{#1}{5_12}}{\specdiag[./KagomeLattice/Diag_5_12]{229}}{}%
\ifthenelse{\equal{#1}{5_13}}{\specdiag[./KagomeLattice/Diag_5_13]{191.5}}{}%
\ifthenelse{\equal{#1}{5_14}}{\specdiag[./KagomeLattice/Diag_5_14]{179}}{}%
\ifthenelse{\equal{#1}{5_15}}{\specdiag[./KagomeLattice/Diag_5_15]{204}}{}%
\ifthenelse{\equal{#1}{5_16}}{\specdiag[./KagomeLattice/Diag_5_16]{266.5}}{}%
\ifthenelse{\equal{#1}{6_1}}{\specdiag[./KagomeLattice/Diag_6_1]{229}}{}%
\ifthenelse{\equal{#1}{6_2}}{\specdiag[./KagomeLattice/Diag_6_2]{229}}{}%
\ifthenelse{\equal{#1}{6_3}}{\specdiag[./KagomeLattice/Diag_6_3]{241.5}}{}%
\ifthenelse{\equal{#1}{6_4}}{\specdiag[./KagomeLattice/Diag_6_4]{279}}{}%
\ifthenelse{\equal{#1}{6_5}}{\specdiag[./KagomeLattice/Diag_6_5]{279}}{}%
\ifthenelse{\equal{#1}{6_6}}{\specdiag[./KagomeLattice/Diag_6_6]{229}}{}%
\ifthenelse{\equal{#1}{6_7}}{\specdiag[./KagomeLattice/Diag_6_7]{241.5}}{}%
\ifthenelse{\equal{#1}{6_8}}{\specdiag[./KagomeLattice/Diag_6_8]{279}}{}%
\ifthenelse{\equal{#1}{6_9}}{\specdiag[./KagomeLattice/Diag_6_9]{279}}{}%
\ifthenelse{\equal{#1}{6_10}}{\specdiag[./KagomeLattice/Diag_6_10]{229}}{}%
\ifthenelse{\equal{#1}{6_11}}{\specdiag[./KagomeLattice/Diag_6_11]{216.5}}{}%
\ifthenelse{\equal{#1}{6_12}}{\specdiag[./KagomeLattice/Diag_6_12]{254}}{}%
\ifthenelse{\equal{#1}{6_13}}{\specdiag[./KagomeLattice/Diag_6_13]{266.5}}{}%
\ifthenelse{\equal{#1}{6_14}}{\specdiag[./KagomeLattice/Diag_6_14]{266.5}}{}%
\ifthenelse{\equal{#1}{6_15}}{\specdiag[./KagomeLattice/Diag_6_15]{216.5}}{}%
\ifthenelse{\equal{#1}{6_16}}{\specdiag[./KagomeLattice/Diag_6_16]{254}}{}%
\ifthenelse{\equal{#1}{6_17}}{\specdiag[./KagomeLattice/Diag_6_17]{254}}{}%
\ifthenelse{\equal{#1}{6_18}}{\specdiag[./KagomeLattice/Diag_6_18]{254}}{}%
\ifthenelse{\equal{#1}{6_19}}{\specdiag[./KagomeLattice/Diag_6_19]{216.5}}{}%
\ifthenelse{\equal{#1}{6_20}}{\specdiag[./KagomeLattice/Diag_6_20]{254}}{}%
\ifthenelse{\equal{#1}{6_21}}{\specdiag[./KagomeLattice/Diag_6_21]{266.5}}{}%
\ifthenelse{\equal{#1}{6_22}}{\specdiag[./KagomeLattice/Diag_6_22]{266.5}}{}%
\ifthenelse{\equal{#1}{6_23}}{\specdiag[./KagomeLattice/Diag_6_23]{216.5}}{}%
\ifthenelse{\equal{#1}{6_24}}{\specdiag[./KagomeLattice/Diag_6_24]{254}}{}%
\ifthenelse{\equal{#1}{6_25}}{\specdiag[./KagomeLattice/Diag_6_25]{254}}{}%
\ifthenelse{\equal{#1}{6_26}}{\specdiag[./KagomeLattice/Diag_6_26]{254}}{}%
\ifthenelse{\equal{#1}{6_27}}{\specdiag[./KagomeLattice/Diag_6_27]{241.5}}{}%
\ifthenelse{\equal{#1}{6_28}}{\specdiag[./KagomeLattice/Diag_6_28]{241.5}}{}%
\ifthenelse{\equal{#1}{6_29}}{\specdiag[./KagomeLattice/Diag_6_29]{354}}{}%
\ifthenelse{\equal{#1}{6_30}}{\specdiag[./KagomeLattice/Diag_6_30]{304}}{}%
\ifthenelse{\equal{#1}{6_31}}{\specdiag[./KagomeLattice/Diag_6_31]{329}}{}%
\ifthenelse{\equal{#1}{6_32}}{\specdiag[./KagomeLattice/Diag_6_32]{304}}{}%
\ifthenelse{\equal{#1}{6_33}}{\specdiag[./KagomeLattice/Diag_6_33]{279}}{}%
\ifthenelse{\equal{#1}{6_34}}{\specdiag[./KagomeLattice/Diag_6_34]{254}}{}%
\ifthenelse{\equal{#1}{6_35}}{\specdiag[./KagomeLattice/Diag_6_35]{254}}{}%
\ifthenelse{\equal{#1}{6_36}}{\specdiag[./KagomeLattice/Diag_6_36]{304}}{}%
\ifthenelse{\equal{#1}{6_37}}{\specdiag[./KagomeLattice/Diag_6_37]{316.5}}{}%
\ifthenelse{\equal{#1}{6_38}}{\specdiag[./KagomeLattice/Diag_6_38]{291.5}}{}%
\ifthenelse{\equal{#1}{7_1}}{\specdiag[./KagomeLattice/Diag_7_1]{241.5}}{}%
\ifthenelse{\equal{#1}{7_2}}{\specdiag[./KagomeLattice/Diag_7_2]{279}}{}%
\ifthenelse{\equal{#1}{7_3}}{\specdiag[./KagomeLattice/Diag_7_3]{291.5}}{}%
\ifthenelse{\equal{#1}{7_4}}{\specdiag[./KagomeLattice/Diag_7_4]{291.5}}{}%
\ifthenelse{\equal{#1}{7_5}}{\specdiag[./KagomeLattice/Diag_7_5]{241.5}}{}%
\ifthenelse{\equal{#1}{7_6}}{\specdiag[./KagomeLattice/Diag_7_6]{279}}{}%
\ifthenelse{\equal{#1}{7_7}}{\specdiag[./KagomeLattice/Diag_7_7]{279}}{}%
\ifthenelse{\equal{#1}{7_8}}{\specdiag[./KagomeLattice/Diag_7_8]{279}}{}%
\ifthenelse{\equal{#1}{7_9}}{\specdiag[./KagomeLattice/Diag_7_9]{291.5}}{}%
\ifthenelse{\equal{#1}{7_10}}{\specdiag[./KagomeLattice/Diag_7_10]{291.5}}{}%
\ifthenelse{\equal{#1}{7_11}}{\specdiag[./KagomeLattice/Diag_7_11]{241.5}}{}%
\ifthenelse{\equal{#1}{7_12}}{\specdiag[./KagomeLattice/Diag_7_12]{279}}{}%
\ifthenelse{\equal{#1}{7_13}}{\specdiag[./KagomeLattice/Diag_7_13]{279}}{}%
\ifthenelse{\equal{#1}{7_14}}{\specdiag[./KagomeLattice/Diag_7_14]{279}}{}%
\ifthenelse{\equal{#1}{7_15}}{\specdiag[./KagomeLattice/Diag_7_15]{291.5}}{}%
\ifthenelse{\equal{#1}{7_16}}{\specdiag[./KagomeLattice/Diag_7_16]{291.5}}{}%
\ifthenelse{\equal{#1}{7_17}}{\specdiag[./KagomeLattice/Diag_7_17]{291.5}}{}%
\ifthenelse{\equal{#1}{7_18}}{\specdiag[./KagomeLattice/Diag_7_18]{279}}{}%
\ifthenelse{\equal{#1}{7_19}}{\specdiag[./KagomeLattice/Diag_7_19]{291.5}}{}%
\ifthenelse{\equal{#1}{7_20}}{\specdiag[./KagomeLattice/Diag_7_20]{291.5}}{}%
\ifthenelse{\equal{#1}{7_21}}{\specdiag[./KagomeLattice/Diag_7_21]{291.5}}{}%
\ifthenelse{\equal{#1}{7_22}}{\specdiag[./KagomeLattice/Diag_7_22]{279}}{}%
\ifthenelse{\equal{#1}{7_23}}{\specdiag[./KagomeLattice/Diag_7_23]{266.5}}{}%
\ifthenelse{\equal{#1}{7_24}}{\specdiag[./KagomeLattice/Diag_7_24]{266.5}}{}%
\ifthenelse{\equal{#1}{7_25}}{\specdiag[./KagomeLattice/Diag_7_25]{266.5}}{}%
\ifthenelse{\equal{#1}{7_26}}{\specdiag[./KagomeLattice/Diag_7_26]{266.5}}{}%
\ifthenelse{\equal{#1}{7_27}}{\specdiag[./KagomeLattice/Diag_7_27]{266.5}}{}%
\ifthenelse{\equal{#1}{7_28}}{\specdiag[./KagomeLattice/Diag_7_28]{266.5}}{}%
\ifthenelse{\equal{#1}{7_29}}{\specdiag[./KagomeLattice/Diag_7_29]{254}}{}%
\ifthenelse{\equal{#1}{7_30}}{\specdiag[./KagomeLattice/Diag_7_30]{254}}{}%
\ifthenelse{\equal{#1}{7_31}}{\specdiag[./KagomeLattice/Diag_7_31]{279}}{}%
\ifthenelse{\equal{#1}{7_32}}{\specdiag[./KagomeLattice/Diag_7_32]{279}}{}%
\ifthenelse{\equal{#1}{7_33}}{\specdiag[./KagomeLattice/Diag_7_33]{279}}{}%
\ifthenelse{\equal{#1}{7_34}}{\specdiag[./KagomeLattice/Diag_7_34]{254}}{}%
\ifthenelse{\equal{#1}{7_35}}{\specdiag[./KagomeLattice/Diag_7_35]{254}}{}%
\ifthenelse{\equal{#1}{7_36}}{\specdiag[./KagomeLattice/Diag_7_36]{354}}{}%
\ifthenelse{\equal{#1}{7_37}}{\specdiag[./KagomeLattice/Diag_7_37]{379}}{}%
\ifthenelse{\equal{#1}{7_38}}{\specdiag[./KagomeLattice/Diag_7_38]{354}}{}%
\ifthenelse{\equal{#1}{7_39}}{\specdiag[./KagomeLattice/Diag_7_39]{266.5}}{}%
\ifthenelse{\equal{#1}{7_40}}{\specdiag[./KagomeLattice/Diag_7_40]{254}}{}%
\ifthenelse{\equal{#1}{7_41}}{\specdiag[./KagomeLattice/Diag_7_41]{266.5}}{}%
\ifthenelse{\equal{#1}{7_42}}{\specdiag[./KagomeLattice/Diag_7_42]{254}}{}%
\ifthenelse{\equal{#1}{7_43}}{\specdiag[./KagomeLattice/Diag_7_43]{279}}{}%
\ifthenelse{\equal{#1}{7_44}}{\specdiag[./KagomeLattice/Diag_7_44]{354}}{}%
\ifthenelse{\equal{#1}{7_45}}{\specdiag[./KagomeLattice/Diag_7_45]{304}}{}%
\ifthenelse{\equal{#1}{7_46}}{\specdiag[./KagomeLattice/Diag_7_46]{304}}{}%
\ifthenelse{\equal{#1}{7_47}}{\specdiag[./KagomeLattice/Diag_7_47]{366.5}}{}%
\ifthenelse{\equal{#1}{7_48}}{\specdiag[./KagomeLattice/Diag_7_48]{316.5}}{}%
\ifthenelse{\equal{#1}{7_49}}{\specdiag[./KagomeLattice/Diag_7_49]{316.5}}{}%
\ifthenelse{\equal{#1}{7_50}}{\specdiag[./KagomeLattice/Diag_7_50]{341.5}}{}%
\ifthenelse{\equal{#1}{7_51}}{\specdiag[./KagomeLattice/Diag_7_51]{291.5}}{}%
\ifthenelse{\equal{#1}{7_52}}{\specdiag[./KagomeLattice/Diag_7_52]{291.5}}{}%
\ifthenelse{\equal{#1}{p_0}}{\specdiag[./KagomeLattice/Diag_p_0]{154}}{}%
\ifthenelse{\equal{#1}{p_1}}{\specdiag[./KagomeLattice/Diag_p_1]{291.5}}{}%
\ifthenelse{\equal{#1}{p_2}}{\specdiag[./KagomeLattice/Diag_p_2]{291.5}}{}%
\ifthenelse{\equal{#1}{p_3}}{\specdiag[./KagomeLattice/Diag_p_3]{266.5}}{}%
\ifthenelse{\equal{#1}{p_4}}{\specdiag[./KagomeLattice/Diag_p_4]{154}}{}%
}%

\newcommand{\sq}[1]%
{%
\ifthenelse{\equal{#1}{1_1}}{\specdiag[./SquareLattice/Diag_1_1]{79}}{}%
\ifthenelse{\equal{#1}{1_2}}{\specdiag[./SquareLattice/Diag_1_2]{79}}{}%
\ifthenelse{\equal{#1}{2_1}}{\specdiag[./SquareLattice/Diag_2_1]{129}}{}%
\ifthenelse{\equal{#1}{2_2}}{\specdiag[./SquareLattice/Diag_2_2]{104}}{}%
\ifthenelse{\equal{#1}{2_3}}{\specdiag[./SquareLattice/Diag_2_3]{129}}{}%
\ifthenelse{\equal{#1}{2_4}}{\specdiag[./SquareLattice/Diag_2_4]{104}}{}%
\ifthenelse{\equal{#1}{2_5}}{\specdiag[./SquareLattice/Diag_2_5]{129}}{}%
\ifthenelse{\equal{#1}{2_6}}{\specdiag[./SquareLattice/Diag_2_6]{104}}{}%
\ifthenelse{\equal{#1}{2_7}}{\specdiag[./SquareLattice/Diag_2_7]{104}}{}%
\ifthenelse{\equal{#1}{2_8}}{\specdiag[./SquareLattice/Diag_2_8]{104}}{}%
\ifthenelse{\equal{#1}{3_1}}{\specdiag[./SquareLattice/Diag_3_1]{179}}{}%
\ifthenelse{\equal{#1}{3_2}}{\specdiag[./SquareLattice/Diag_3_2]{154}}{}%
\ifthenelse{\equal{#1}{3_3}}{\specdiag[./SquareLattice/Diag_3_3]{104}}{}%
\ifthenelse{\equal{#1}{3_4}}{\specdiag[./SquareLattice/Diag_3_4]{129}}{}%
\ifthenelse{\equal{#1}{3_5}}{\specdiag[./SquareLattice/Diag_3_5]{179}}{}%
\ifthenelse{\equal{#1}{3_6}}{\specdiag[./SquareLattice/Diag_3_6]{154}}{}%
\ifthenelse{\equal{#1}{3_7}}{\specdiag[./SquareLattice/Diag_3_7]{154}}{}%
\ifthenelse{\equal{#1}{3_8}}{\specdiag[./SquareLattice/Diag_3_8]{104}}{}%
\ifthenelse{\equal{#1}{3_9}}{\specdiag[./SquareLattice/Diag_3_9]{129}}{}%
\ifthenelse{\equal{#1}{3_10}}{\specdiag[./SquareLattice/Diag_3_10]{129}}{}%
\ifthenelse{\equal{#1}{3_11}}{\specdiag[./SquareLattice/Diag_3_11]{129}}{}%
\ifthenelse{\equal{#1}{3_12}}{\specdiag[./SquareLattice/Diag_3_12]{129}}{}%
\ifthenelse{\equal{#1}{4_1}}{\specdiag[./SquareLattice/Diag_4_1]{229}}{}%
\ifthenelse{\equal{#1}{4_2}}{\specdiag[./SquareLattice/Diag_4_2]{204}}{}%
\ifthenelse{\equal{#1}{4_3}}{\specdiag[./SquareLattice/Diag_4_3]{179}}{}%
\ifthenelse{\equal{#1}{4_4}}{\specdiag[./SquareLattice/Diag_4_4]{179}}{}%
\ifthenelse{\equal{#1}{4_5}}{\specdiag[./SquareLattice/Diag_4_5]{154}}{}%
\ifthenelse{\equal{#1}{4_6}}{\specdiag[./SquareLattice/Diag_4_6]{154}}{}%
\ifthenelse{\equal{#1}{4_7}}{\specdiag[./SquareLattice/Diag_4_7]{129}}{}%
\ifthenelse{\equal{#1}{4_8}}{\specdiag[./SquareLattice/Diag_4_8]{129}}{}%
\ifthenelse{\equal{#1}{4_9}}{\specdiag[./SquareLattice/Diag_4_9]{129}}{}%
\ifthenelse{\equal{#1}{4_10}}{\specdiag[./SquareLattice/Diag_4_10]{204}}{}%
\ifthenelse{\equal{#1}{4_11}}{\specdiag[./SquareLattice/Diag_4_11]{179}}{}%
\ifthenelse{\equal{#1}{4_12}}{\specdiag[./SquareLattice/Diag_4_12]{179}}{}%
\ifthenelse{\equal{#1}{4_13}}{\specdiag[./SquareLattice/Diag_4_13]{179}}{}%
\ifthenelse{\equal{#1}{4_14}}{\specdiag[./SquareLattice/Diag_4_14]{154}}{}%
}%

\newcommand{\hon}[1]%
{%
\ifthenelse{\equal{#1}{2_1}}{\specdiag[./HoneycombLattice/Diag_2_1]{154}}{}%
\ifthenelse{\equal{#1}{2_2}}{\specdiag[./HoneycombLattice/Diag_2_2]{154}}{}%
\ifthenelse{\equal{#1}{4_1}}{\specdiag[./HoneycombLattice/Diag_4_1]{229}}{}%
\ifthenelse{\equal{#1}{4_2}}{\specdiag[./HoneycombLattice/Diag_4_2]{191.5}}{}%
\ifthenelse{\equal{#1}{4_3}}{\specdiag[./HoneycombLattice/Diag_4_3]{229}}{}%
\ifthenelse{\equal{#1}{4_4}}{\specdiag[./HoneycombLattice/Diag_4_4]{229}}{}%
\ifthenelse{\equal{#1}{4_5}}{\specdiag[./HoneycombLattice/Diag_4_5]{191.5}}{}%
\ifthenelse{\equal{#1}{4_6}}{\specdiag[./HoneycombLattice/Diag_4_6]{191.5}}{}%
\ifthenelse{\equal{#1}{4_7}}{\specdiag[./HoneycombLattice/Diag_4_7]{191.5}}{}%
}%

\def\fileversion{v1.2a}
\def\filedate{2007/11/23}
\makeatletter

\newbox\swb@xone
\newbox\swb@xtwo
\newbox\swb@xthree
\newbox\swb@xfour
\newdimen\swdimen@ne
\newdimen\swdimentw@

\newcommand{\acontraction}[5][1ex]{%
  \mathchoice
    {\acontraction@\displaystyle{#2}{#3}{#4}{#5}{#1}}%
    {\acontraction@\textstyle{#2}{#3}{#4}{#5}{#1}}%
    {\acontraction@\scriptstyle{#2}{#3}{#4}{#5}{#1}}%
    {\acontraction@\scriptscriptstyle{#2}{#3}{#4}{#5}{#1}}}%
\newcommand{\acontraction@}[6]{%
  \setbox\swb@xone=\hbox{${}#1{}#2{}$}%
  \setbox\swb@xtwo=\hbox{${}#1{}#3{}$}%
  \setbox\swb@xthree=\hbox{${}#1{}#4{}$}%
  \setbox\swb@xfour=\hbox{${}#1{}#5{}$}%
  \swdimen@ne=\wd\swb@xtwo%
  \advance\swdimen@ne by \wd\swb@xfour%
  \divide\swdimen@ne by 2%
  \advance\swdimen@ne by \wd\swb@xthree%
  \vbox{%
    \hbox to 0pt{%
      \kern \wd\swb@xone%
      \kern 0.5\wd\swb@xtwo%
      \acontraction@@{\swdimen@ne}{#6}%
      \hss}%
    \vskip 0.5ex
    \vskip\ht\swb@xtwo}}

\newcommand{\acontraction@@}[3][0.05em]{%
  \hbox{%
    \vrule width #1 height 0pt depth #3%
    \vrule width #2 height 0pt depth #1%
    \vrule width #1 height 0pt depth #3%
    \relax}}

\newcommand{\bcontraction}[5][1ex]{%
  \mathchoice
    {\bcontraction@\displaystyle{#2}{#3}{#4}{#5}{#1}}%
    {\bcontraction@\textstyle{#2}{#3}{#4}{#5}{#1}}%
    {\bcontraction@\scriptstyle{#2}{#3}{#4}{#5}{#1}}%
    {\bcontraction@\scriptscriptstyle{#2}{#3}{#4}{#5}{#1}}}%
\newcommand{\bcontraction@}[6]{%
  \setbox\swb@xone=\hbox{${}#1{}#2{}$}%
  \setbox\swb@xtwo=\hbox{${}#1{}#3{}$}%
  \setbox\swb@xthree=\hbox{${}#1{}#4{}$}%
  \setbox\swb@xfour=\hbox{${}#1{}#5{}$}%
  \swdimen@ne=\wd\swb@xtwo%
  \advance\swdimen@ne by \wd\swb@xfour%
  \divide\swdimen@ne by 2%
  \advance\swdimen@ne by \wd\swb@xthree%
  \lower 0.5ex \vbox{%
    \hbox to 0pt{%
      \kern \wd\swb@xone%
      \kern 0.5\wd\swb@xtwo%
      \bcontraction@@{\swdimen@ne}{#6}%
      \hss}%
    }}

\newcommand{\bcontraction@@}[3][0.05em]{%
  \hbox{%
    \swdimentw@=#3
    \advance\swdimentw@ by -#1
    \vrule width #1 height 0pt depth #3%
    \lower\swdimentw@\hbox{\vrule width #2 height 0pt depth #1}%
    \vrule width #1 height 0pt depth #3%
    \relax}}

\makeatother

\newcommand{\ket}[1]{\vert{#1} \rangle}

\newcommand{\1}[1][]{\hat{1}^{#1}}
\newcommand*{\vcenteredhbox}[1]{\begingroup
\setbox0=\hbox{#1}\parbox{\wd0}{\box0}\endgroup}
\newcommand{\state}[2][]{\vcenteredhbox{\includegraphics[width=#2]{#1}}}

\def\horparallel{ \lower.5ex\hbox{ \includegraphics[width=2ex]{fig-hor.pdf}}\,\, }
\def\vertparallel{ \lower.5ex\hbox{ \includegraphics[width=2ex]{fig-vert.pdf}}\,\, }

\graphicspath{{Figs/}{SmallFigs/}}

\begin{document}

\title{Engineering SU(2) invariant spin models to mimic quantum dimer physics on the square lattice}

\author{M. Mambrini}
\email{mambrini@irsamc.ups-tlse.fr}
\author{S. Capponi}
\email{capponi@irsamc.ups-tlse.fr}
\author{F. Alet}
\email{alet@irsamc.ups-tlse.fr}
\affiliation{Laboratoire de Physique Th\'eorique, Universit\'e de Toulouse and CNRS, UPS (IRSAMC), F-31062 Toulouse, France}

\date{\today}

\begin{abstract}

We consider the spin-$1/2$ hamiltonians proposed by Cano and Fendley [J. Cano and P. Fendley, Phys. Rev. Lett. {\bf 105}, 067205 (2010)] which were built to promote the well-known Rokshar-Kivelson (RK) point of quantum dimer models to spin-$1/2$ wavefunctions. We first show that these models, besides the exact degeneracy of RK point, support gapless spinless excitations as well as a spin gap in the thermodynamic limit, signatures of an unusual spin liquid. We then extend the original construction to create a continuous family of SU(2) invariant spin models that reproduces the phase diagram of the quantum dimer model, and in particular show explicit evidences for existence of columnar and staggered phases. The original models thus appear as multicritical points in an extended phase diagram. Our results are based on the use of a combination of numerical exact simulations and analytical mapping to effective generalized quantum dimer models.

\end{abstract}

\pacs{
{75.10.Jm},
{75.40.Mg} 
}

\maketitle

\section*{Introduction}

Construction of low-energy effective models has a long history in physics,
when the original model of interest is too complicated to solve or to be
treated with perturbation theory. This is true for instance in lattice quantum chromodynamics
or in condensed matter physics. When constructing such low-energy effective
models, it is important to remember how the initial degrees of freedom are
transformed, especially if they are naturally coupled to experimental probes.

A canonical example in condensed matter physics is found in the
context of Mott insulators where the charges are frozen and the spins become
the relevant degrees of freedom. In that case, the electronic Hubbard model
at half-filling in the large interaction limit can be reduced to the
Heisenberg model
\begin{equation}
\label{eq:HJ}
 H_J = J \sum_{\langle i,j \rangle} {\bf S_i} . {\bf S_j}
\end{equation}
which only involves spin degrees of freedom. The exchange constant $J$ favors
antiferromagnetic order in the ground-state (GS) when positive and the sum runs over nearest neighbors
(NN) spins. Quantum magnets exhibit a large plethora of exotic phenomena,
especially when the exchange interactions are frustrated. There, magnetic
long-range order can be prohibited and two NN spins
naturally tend to lock in a non-magnetic two-sites singlet or valence bond (VB) $(\ket{\uparrow
  \downarrow}-\ket{\downarrow \uparrow})/\sqrt{2}$. 

When the physics is local and non-magnetic such as often found in frustrated
magnets, it has been argued that NNVBs can serve as the
correct emergent degrees of freedom. It is therefore
tempting to derive an effective model in terms of these short-range
objects. However, this is a difficult task as VB states are non-orthogonal.
Building on this intuition, Rokhsar and Kivelson (RK) constructed a model\cite{RK}
allowing quantum dynamics for two-sites short-range objects, which can be
derived from the first terms of an overlap expansion of the Heisenberg
model. The resulting quantum dimer model~(QDM)
\begin{align}
  \nonumber H_{\rm QDM}= \sum \Big[ & -t \left (
	\;\state[sq2]{0.6cm} \;\state[qs1]{0.6cm} + \rm{h.c.}\right) + \\ &
  \quad\quad v \left(  \;\state[sq2]{0.6cm} \;\state[qs2]{0.6cm}  + \; \state[sq1]{0.6cm} \;\state[qs1]{0.6cm} \;\right) \Big]\,,
  \label{eq:HQDM}
\end{align}
(here depicted on a plaquette of the square lattice) involves (NN) {\it quantum dimers},
which are hardcore objects with no internal structure (in contrast with the antisymmetric VB where the spin
degrees of freedom can still be probed). Quite importantly, quantum dimer
states are taken as orthogonal.  An important feature of
Eq.~(\ref{eq:HQDM}) is the existence of a so-called RK point at $v=t$ where
the ground-state is exactly known and is an equal amplitude state in each
ergodicity sector of $H_{\rm QDM}$ (see Appendix~\ref{appendix1}).

Besides their relative simplicity due to the use of an orthogonal basis, there are several practical advantages to use QDMs instead of (frustrated) spin models. They are indeed more tractable analytically (existence of RK points \cite{RK}, some QDMs are even exactly solvable\cite{Misguich2002}) and numerically (the reduced Hilbert space size allows the study of larger systems with exact diagonalization and very often no sign problem appears in quantum Monte Carlo simulations). 
These technical aspects only partly explain the popularity of QDMs, the
other ingredient being the richness of their phase diagrams~\cite{Moessner2002}. Indeed, several
unconventional phase of matters such as $Z_2$ or $U(1)$ liquids or valence
bond crystals (VBC) of different types (columnar, staggered, plaquette or
mixed) have been exhibited in QDMs~\cite{Moessner2001a,Misguich2002,Moessner2003,Sikora2011,Ralko2005}, which are hard to find or
characterize in spin models.

Quite crucially however, the initial spin degrees of freedom are completely
forgotten in QDMs as they all are assumed to pair in NN singlets. They are
yet essential if one wishes to allow magnetic order to compete, or just to
compare with experimental results which probe two-spin (neutron scattering)
or four-spin (Raman scattering) correlations. In QDMs, two-spin correlations
$C_{ij}=\langle {\bf S}_i \cdot {\bf S}_j \rangle$ cannot be easily computed
while four-spin correlators $C_{ijkl}=\langle ({\bf S}_i\cdot {\bf S}_j )
({\bf S}_k\cdot {\bf S}_l) \rangle - C_{ij}.C_{kl}$ are mimicked by (hardcore) dimer-dimer occupancy correlations $\langle n_{ij} n_{kl}  \rangle$ in the sole case where $i,j$ and $k,l$ are NN.

Clarifying the connection between SU(2) spin models and QDM is a long standing question. One interesting route was investigated in Ref.~\onlinecite{Raman2005} using a decoration scheme or more recently using Projected Entangled Pair States (PEPS)\cite{Schuch2012}. In this spirit, a systematic way of deriving QDMs from Heisenberg models was introduced by one of the authors and coworkers\cite{Ralko2009,Poilblanc2010,Schwandt2010}, and the resulting
effective models, dubbed Generalized Quantum Dimer Models (GQDM), involve terms on larger plaquettes. 

In an insightful work~\cite{CF}, Cano and Fendley (CF) took a step in the other direction and introduced two spin-$1/2$ models with local spin projectors that display RK-like GS in the NNVB basis. In this article, we (i) first investigate the low energy properties of the two CF models, and (ii) extend them by constructing and studying a continuous family of spin-$1/2$ models that closely mimics QDM physics. These spin models share useful properties with QDMs (exactly known GS, good variational approach in a reduced Hilbert space), while retaining the true spin degrees of freedom (allowing spin order to set in and computations of spin correlations). Using an effective model approach as well as exact diagonalizations on both the original spin-$1/2$ model and its projection on the NNVB basis, we show that the obtained phase diagram is in correspondence with the one of QDM on the square lattice, including explicit signatures of VBC as well of existence of a RK-like multicritical point.

The outline of this paper is as follows. We first briefly recall in Sec.~\ref{sec:model} the most important aspects of the CF construction and introduce the family of spin models studied along the paper. We then consider in Sec.~\ref{sec:pureCF} both CF models, and show using exact diagonalizations up to 40 spins and diagonalizations in the variational NNVB basis up to 50 spins that they indeed possess RK-like ground-states but also have a low-energy excitation structure similar to the RK point of the QDM, in particular gapless singlet excitations. In the second part of this manuscript, we extend in Sec.~\ref{sec:mixingCF} the CF construction to a continuous family of spin models and establish a correspondence between the SU(2)-invariant spin-$1/2$ models and the QDM phase diagrams for arbitrary $v/ \vert t \vert$ values. This is performed using numerical simulations as well as analytical arguments based on a GQDM mapping of the spin hamiltonian. We finally close with a discussion in Sec.~\ref{sec:conc}. Two appendices provide details on the exact degeneracy of square-lattice QDM models at their RK point (App.~\ref{appendix1}), as well as on the GQDM mapping of CF models and their extensions (App.~\ref{appendix2}).

\section{Cano-Fendley Models and their generalization}\label{sec:model}

We recall here the main ingredients of the Cano-Fendley~\cite{CF} construction to obtain spin-$1/2$ models with ground-state(s) that are equal-amplitude superpositions of NNVB coverings, here on the square lattice. 
The Cano-Fendley models,
\begin{equation}
\label{eq:models}
\hat{\cal H}^\mu = \hat{{\cal H}}_1 + \hat{{\cal H}}^\mu_2 \qquad\mu=0,1
\end{equation}
contain two terms:
\begin{equation}
\label{eq:klein}
\hat{{\cal H}}_1 = \sum_{\includegraphics[width=0.2cm]{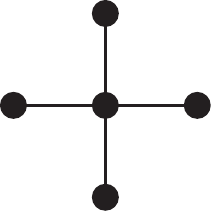}} \hat{P}_{\includegraphics[width=0.2cm]{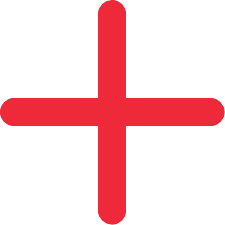}}^{S=5/2},
\end{equation}
known as a Klein term~\cite{Klein}, 
and 
\begin{align}
\label{eq:cfendley}
\hat{{\cal H}}^\mu_2 &= \sum_{\includegraphics[width=0.4cm]{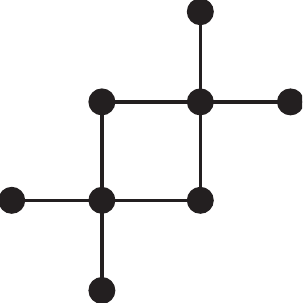}} \hat{P}_{\includegraphics[width=0.4cm]{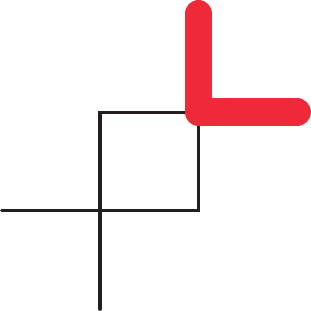}}^{S=3/2} \hat{P}_{\includegraphics[width=0.4cm]{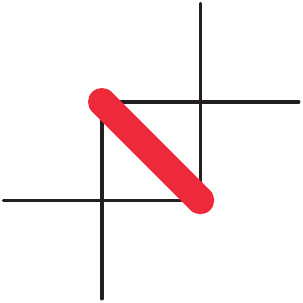}}^{S=\mu} \hat{P}_{\includegraphics[width=0.4cm]{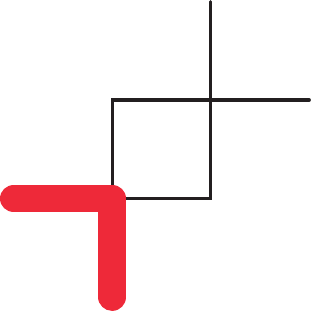}}^{S=3/2} \nonumber \\
&+ {\sum_{\includegraphics[width=0.4cm,angle=90]{CF_gen}}} \hat{P}_{\includegraphics[angle=90,width=0.4cm]{CF_1}}^{S=3/2} \hat{P}_{\includegraphics[angle=90,width=0.4cm]{CF_3}}^{S=\mu} \hat{P}_{\includegraphics[angle=90,width=0.4cm]{CF_2}}^{S=3/2},
\end{align}
where the sums runs over all the square plaquettes $p$ of the lattice and $\hat{P}^S_c$ stands for the projection operator on the total spin $S=\mu$ (with $\mu=0$ or $1$) sector on the cluster $c$. We consider here square lattices with periodic boundary conditions.

For convenience, we recall that these projectors can be easily reexpressed in terms of spins operators and thus Eq.~(\ref{eq:models}) reinterpreted as multispin interactions :
\begin{align*}
\label{eq:explicitprojectors}
\hat{P}_{\includegraphics[width=0.2cm]{CF_gen2}}^{S=5/2} &= \frac{1}{10} \Big (\sum_{(i,j) \in \; \vcenter{\hbox{\includegraphics[width=0.3cm]{CF_gen3}}}} 
\mathbf{S}_i.\mathbf{S}_j \Big ) \Big (\frac{3}{2}+\sum_{(i,j) \in \; \vcenter{\hbox{\includegraphics[width=0.3cm]{CF_gen3}}}} 
\mathbf{S}_i.\mathbf{S}_j \Big )\\
\hat{P}_{\includegraphics[width=0.4cm]{CF_1}}^{S=3/2} &= \frac{2}{3} \Big (\frac{3}{4}+\sum_{(i,j) \in \; \vcenter{\hbox{\includegraphics[width=0.4cm]{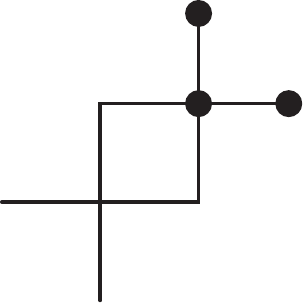}}}} \mathbf{S}_i.\mathbf{S}_j \Big ) \\
\hat{P}_{\includegraphics[width=0.4cm]{CF_3}}^{S=0} &= \frac{1}{4} - \mathbf{S}_i.\mathbf{S}_j \\
\hat{P}_{\includegraphics[width=0.4cm]{CF_3}}^{S=1} &=\frac{3}{4} + \mathbf{S}_i.\mathbf{S}_j.
\end{align*}

The roles of the two terms (\ref{eq:klein}) and (\ref{eq:cfendley}) are different. The Klein term $\hat{{\cal H}}_1$ is a projection term designed to annihilate any NNVB state. Since this operator is positive, this ensure that all NNVB
are located in its groundstate manifold. Note that this manifold also contains non NNVB states (e.g. VB states with aligned VB on the diagonal of plaquettes) -- at this stage, the ground-state manifold of $\hat{{\cal H}}_1$ is therefore largely degenerate. This term is gapped to other singlet excitations, and this gap is expected to be robust when introducing $\hat{{\cal H}}^\mu_2$. 

$\hat{{\cal H}}^\mu_2$ is designed to introduce dynamics in the GS manifold of $\hat{{\cal H}}_1$, and therefore to lift its large degeneracy. The form of $\hat{{\cal H}}^\mu_2$ will be such as to select the equal-amplitude superposition of NNVBs as the ground-state of $\hat{{\cal H}}_1+\hat{{\cal H}}^\mu_2$. To see this, let us detail the dynamics introduced by $\hat{{\cal H}}^\mu_2$ on a given NNVB state $| \psi \rangle$~:

\begin{enumerate}
\item Case 1: $p$ contains $0$ or $1$ dimer in $| \psi \rangle$ (non-flippable plaquette). This implies that at least one dimer is hosted by a corner sharing neighboring plaquette. In that case either $\hat{P}_{\includegraphics[width=0.4cm]{CF_1}}^{S=3/2}$, $\hat{P}_{\includegraphics[width=0.5cm]{CF_2}}^{S=3/2}$, $\hat{P}_{\includegraphics[angle=90,width=0.4cm]{CF_1}}^{S=3/2}$ or $\hat{P}_{\includegraphics[angle=90,width=0.4cm]{CF_2}}^{S=3/2}$ annihilates the state. 
\item Case 2: $p$ contains $2$ dimers (flippable plaquette). We note that
\begin{equation}
\hat{P}_{\includegraphics[width=0.4cm]{CF_3}}^{S=\mu}=\frac{1}{2} \left ( (2\mu-1) \hat{\Pi}_{\includegraphics[width=0.4cm]{CF_3}} +\1 \right ),
\end{equation}
where $\hat{\Pi}_{ij}$ is the permutation operator  on sites $i$ and $j$.
It is easy to show that :
\begin{align}
\hat{\Pi}_{\includegraphics[width=0.4cm]{CF_3}}   \state[Flippable1]{0.8cm} &=  \state[Flippable2]{0.8cm}\\
\hat{\Pi}_{\includegraphics[width=0.4cm]{CF_3}}   \state[Flippable2]{0.8cm} &=  \state[Flippable1]{0.8cm}\\
\hat{\Pi}_{\includegraphics[width=0.4cm,angle=90]{CF_3}}  \state[Flippable1]{0.8cm} &= \state[Flippable2]{0.8cm}\\
\hat{\Pi}_{\includegraphics[width=0.4cm,angle=90]{CF_3}}  \state[Flippable2]{0.8cm} &= \state[Flippable1]{0.8cm}.
\end{align}

Hence,
\begin{align}
\label{eq:dyn1}
\left ( \hat{P}_{\includegraphics[width=0.4cm]{CF_3}}^{S=\mu}  + \hat{P}_{\includegraphics[width=0.4cm,angle=90]{CF_3}}^{S=\mu} \right) \state[Flippable1]{0.8cm} &=  (2\mu-1) \; \state[Flippable2]{0.8cm} + \; \state[Flippable1]{0.8cm} \\
\label{eq:dyn2}
\left ( \hat{P}_{\includegraphics[width=0.4cm]{CF_3}}^{S=\mu} + \hat{P}_{\includegraphics[width=0.4cm,angle=90]{CF_3}}^{S=\mu} \right ) \state[Flippable2]{0.8cm} &= (2\mu-1) \; \state[Flippable1]{0.8cm} +  \; \state[Flippable2]{0.8cm}.
\end{align}
\end{enumerate}

At this stage, it seems from Eq.~(\ref{eq:dyn1})-(\ref{eq:dyn2}) that $\hat{{\cal H}}^\mu_2$ perfectly mimicks the QDM Eq.~(\ref{eq:HQDM}) in
the variational NNVB basis with $v=1$ and $t=1$ (respectively $t=-1$ ) for $\mu=0$ (respectively $\mu=1$). 
However, it is important to remark that the states involved in the r.h.s. of Eq.~(\ref{eq:dyn1})-(\ref{eq:dyn2}) are not invariant under the application of $\hat{P}_{\includegraphics[width=0.4cm]{CF_1}}^{S=3/2}$ or $\hat{P}_{\includegraphics[width=0.4cm]{CF_2}}^{S=3/2}$ and these operators induce further reconfigurations of NNVB states outside the flippable plaquette. This will lead to subtle differences between the two models that will be discussed in the following, together with the precise correspondence between CF models and their QDM counterparts.

Note as well that on the square lattice the
sign of $t$ is irrelevant for the QDM as it can be gauged away. Therefore, both CF models are designed to have the equal-amplitude states in the NNVB {\it non-orthogonal} basis~\cite{Sutherland,Albuquerque,Tang} as zero-energy GS. Similarly to the situation of the RK-point of the QDM, all equal-amplitude states in each ergodicity sector are valid GS, therefore the minimal degeneracy of the CF models is equal to the number of ergodicity / topological sectors for NNVB / dimer coverings of the square lattice (see Appendix~\ref{appendix1}). Note that we used the term 'minimal' as other GS may arise due to the Klein term which also annihilates non-NNVBs states on a square lattice with periodic boundary conditions. In Sec.~\ref{sec:pureCF} we present numerical evidence, using exact diagonalizations both in the full $S_z$ basis (allowing to study magnetic and non-magnetic states) and in the variational NNVB basis (expected to be relevant there), that the models proposed by Cano and Fendley  are indeed realizations of SU(2)-invariant spin-$1/2$ models that not only share the GS properties but also the low-energy structure with the QDM at the RK point.

In Sec.~\ref{sec:mixingCF} of this article, we investigate, using various numerical and analytical methods, a generalization of Cano and Fendley proposal allowing to mix  both CF models $\hat{{\cal H}}^0$ and $\hat{{\cal H}}^1$~:
\begin{equation}
\label{eq:generalmodel}
\hat{\cal H}^{K,\theta} = K \hat{{\cal H}}_1 + \frac{1}{ n ( \theta )} \left ( \cos \theta \; \hat{\cal H}_2^0 + \sin \theta \; \hat{\cal H}_2^1 \right ),
\end{equation}
where the normalization factor $n(\theta) = \vert \cos \theta \vert + \vert \sin \theta \vert$ is introduced for convenience. The two CF models are recovered as $\hat{{\cal H}}^0= \hat{\cal H}^{K=1,\theta=0}$ and  $\hat{{\cal H}}^1= \hat{\cal H}^{K=1,\theta=\pi/2}$. Note that $\hat{\cal H}^{K,\theta}$ also conserves the SU(2) spin symmetry of the CF model, as well as square-lattice spatial symmetries.

The motivation to introduce $\hat{\cal H}^{K,\theta}$ is to see whether this family of spin-$1/2$ models can host phases similar to those observed in the phase diagram of the conceptually simpler QDM on the square lattice (Eq.~\ref{eq:HQDM}), as the mixing angle $\theta$ is tuned.

\section{Cano-Fendley models: $\theta=0$ and $\theta=\pi/2$}\label{sec:pureCF}

\subsection{Prefactor of the Klein term}\label{sec:Klein}

Before going into the detailed numerical study of $\hat{\cal H}^{K,0}$ and $\hat{\cal H}^{K,\pi/2}$, it is important to investigate the influence of the prefactor $K$ of the Klein term $\hat{{\cal H}}_1$ in Eq.~(\ref{eq:generalmodel}) as it plays an important role to isolate the NNVB states from other singlet or triplet states in the low-energy manifold. 

\begin{figure}
\includegraphics[width=0.8\columnwidth]{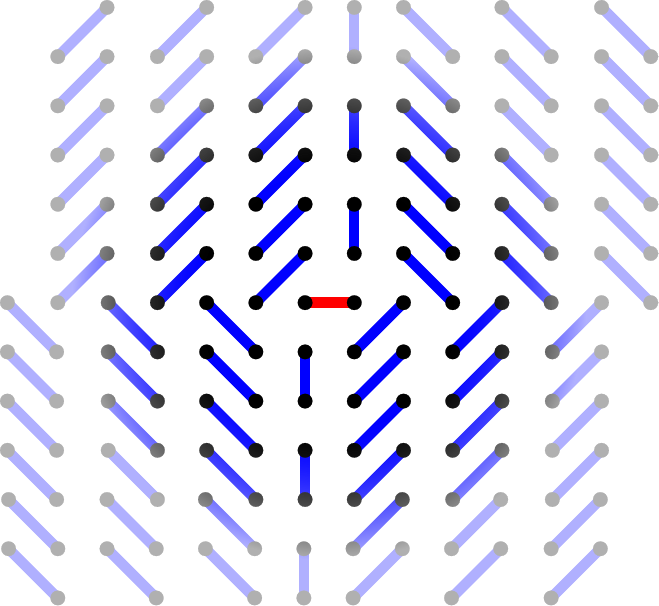}
\caption{(color online) Non-NNVB zero energy singlet (resp. zero energy triplet) eigenstate of the Klein term: the central red bond is to be interpreted as a singlet (resp. a triplet).}
\label{fig:Gapless_singlets_and_triplet}
\end{figure}

First let us consider the Klein term alone. On a square lattice with periodic boundary conditions, the Klein term is known to be \emph{not perfect}~\cite{Chayes1989} in isolating the NNVB manifold, as other singlet states and even triplets also have zero energy\footnote{With open boundary conditions the Klein term may be perfect on the square lattice. We nevertheless restrict ourselves to periodic boundary conditions which are more suitable for numerical analysis.}.  Indeed, it is quite easy to verify that the states depicted in Fig.~\ref{fig:Gapless_singlets_and_triplet} are zero energy non-NNVB singlet or triplet eigenstates of the Klein term, as the total spin of any group of five spins formed by a site and its four nearest neighbors is lower than $3/2$.

Since the Klein term has no spin gap, it is important to show that there exists a regime of value for $K$ \emph{in the presence of} ${\cal H}_2^{\mu}$ for which (i) a sizable spin gap opens in the thermodynamic limit and (ii) the other interactions included in ${\cal H}_2^\mu$ allow NNVB configurations to gain further energy by resonating, and therefore to separate from other zero-energy states of the Klein term. 

\begin{figure}
\includegraphics[width=\columnwidth]{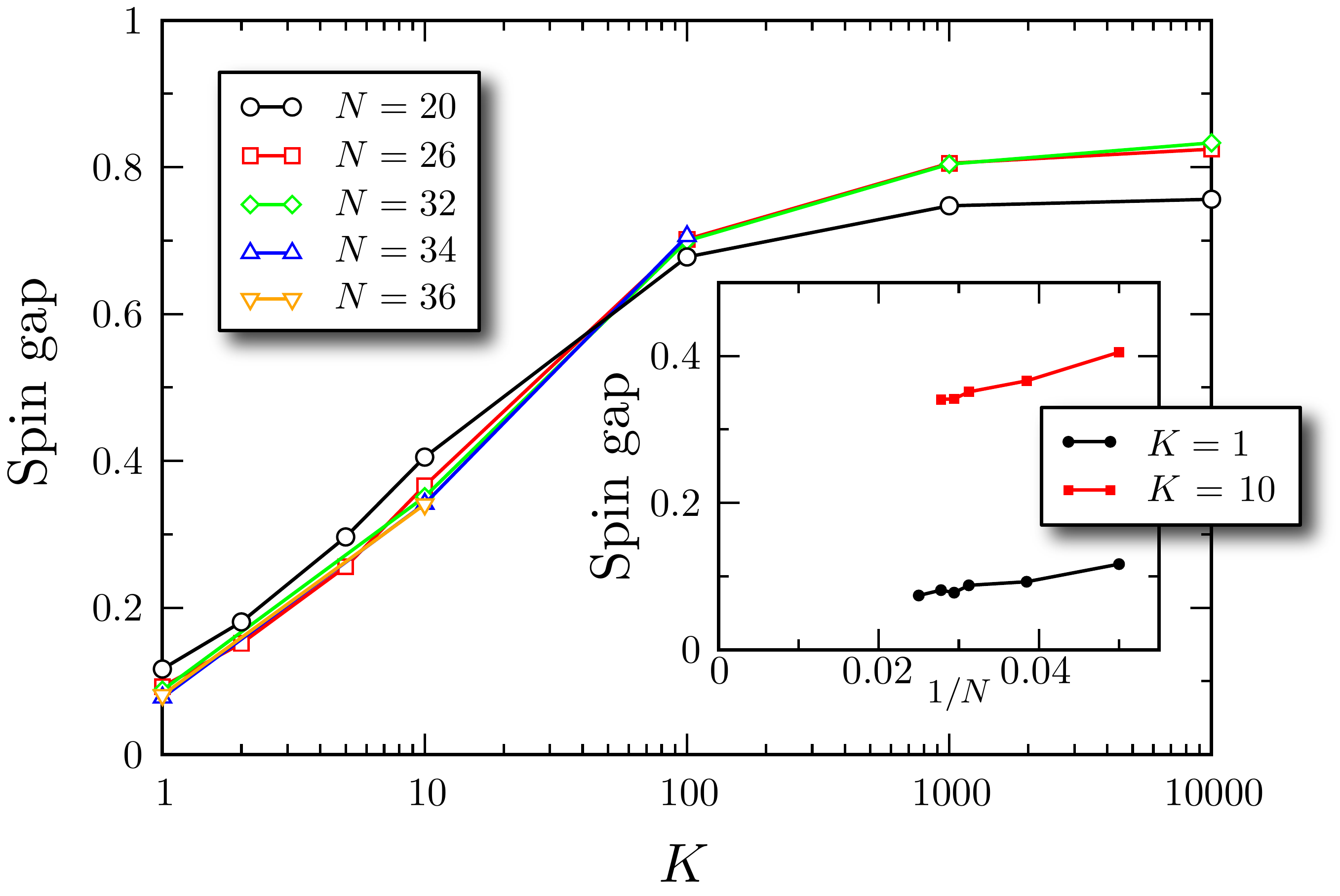}
\caption{(color online) Spin gap versus amplitude $K$ of the Klein term for the ${\cal H}^{K,0}$ model on different square clusters with $N$ sites. Inset: scaling of the spin gap vs $1/N$ for two values of $K$. }
\label{fig:spingap_vs_Klein}
\end{figure}

In the original CF proposal, its amplitude was set to 1, but we find that it is necessary to consider a larger value. We show in Fig.~\ref{fig:spingap_vs_Klein} the evolution of the spin gap vs the amplitude $K$ for the model $\hat{\cal H}^{K,\theta=0}$. The finite-size behavior is not always monotonous, but a finite-size scaling analysis (see inset) indicates a finite spin gap in the thermodynamic limit quite convincingly when $K\geq 10$, while data with $K=1$ are more ambiguous. Since the iterative exact diagonalization techniques that we use for the largest samples become slower when $K$ is too large, we find it practical to fix its value to $K=10$, which will be enough to ensure that all models studied further have a finite spin gap. This value $K=10$ will be implicitly used in the rest of this work. We will show in the next section that it is also large enough to eliminate non-NNVB intruder singlet states from the low energy manifold.

As a final remark, let mention that in the NNVB approach (consisting in a ED of the spin hamiltonian in the subspace spanned by NNVB states) that we use in conjunction with ED in the $S_z$ basis, the two points mentioned above are naturally fulfilled since, by construction, neither triplets nor non-NNVB states enter the computation. In a sense using this approach is equivalent to forcing the Klein term to be perfect. 

\subsection{Ground-state degeneracy and properties of CF models}\label{sec:GS}

Here we discuss the degeneracy of the ground-states for the $\hat{\cal H}^{K,0}$ and $\hat{\cal H}^{K,\pi/2}$ models. All ground-states at these points have an energy which is exactly 0. Since one can build an equal-amplitude ground-state in each ergodicity sector, we expect at least the same number of ground-states as at the RK point of the QDM. In our exact diagonalization computations, two cases need to be distinguished for the computation of the degeneracy of these zero-energy states: (i) we can compute exactly the full spectrum for diagonalizations in the variational NNVB basis as well as in the $S^z$ basis for small enough samples  ($N \leq 20$), (ii) when we cannot compute the full spectrum in the $S^z$ basis even when decomposing it into symmetry sectors, we use Davidson algorithm to compute exactly the (possibly degenerate) lowest-lying states.

We find that the number of zero-energy GS of the CF and QDM models (at the RK point) are {\it exactly} the same (see table \ref{tab:deg}), except for the sample $N=16$ which exhibits 4 spurious ground-states (3
singlets and even a triplet) leading to a degeneracy of 23 instead of 17. 

The NNVB calculation does not take into account (by construction) triplets or long range singlets and does not display the $N=16$ anomaly in the GS degeneracy. This allows us to interpret spurious singlets as a small size artifact: on such a small cluster, long-range singlets cannot exceed 4 lattice spacings and are still sensitive to the 8-spin CF terms, but this anomaly does not persist for larger clusters. The degeneracies of both $\hat{\cal H}^{K,0}$ and $\hat{\cal H}^{K,\pi/2}$ in NNVB calculations are thus matching the ones of the QDM at RK point for all cluster sizes from $N=16$ to $N=50$. We also explicitly checked that each GS is indeed of equal-amplitude type in the NNVB basis, exactly as the QDM GS in the orthogonal quantum dimer basis. The spatial quantum numbers are also identical.

\begin{table}
  \centering
\begin{tabular}{cccccccc}
\hline\hline
N 			& 16 & 20 & 26 & 32 & 36 & 40 & 50 \\
GS deg. & 17 & 13 & 16 & 69 & 41 & 29 & 47 \\
\hline\hline
\end{tabular}
  \caption{GS degeneracies of the QDM at the RK point. Except for $N=16$ where spurious states enter the GS manifold, these degeneracies are identical for the CF models $\hat{\cal H}^{K,0}$ and $\hat{\cal H}^{K,\pi/2}$.}
  \label{tab:deg}
\end{table}

The physical properties of the GS of the CF model and the QDM are slightly different
than the ones of the QDM. First of all, spin correlations can be computed
(opposite to the QDM as discussed in the introduction) and are known since a
long time~\cite{LDA} to be exponentially decaying $C_{ij}\propto
\exp(-|i-j|/\xi)$ with $\xi=1.35$~\cite{Albuquerque,Tang}. More recently, it
was discovered that four-spin correlators $C_{ijkl}$ (with $i,j$ and $k,l$
NN) rather decay algebraically with an exponent $\alpha_{\rm NNVB} \simeq
1.16$~\cite{Albuquerque,Tang}, {\it i.e.} more slowly than in the QDM case
where $\alpha_{\rm QDM}=2$ for the decay of dimer-dimer correlators~\cite{FisherStephenson,Damle2012}. 
These critical four-spin correlations in the ground-state strongly suggest~\cite{Albuquerque,CF} that the CF models are gapless with respect to singlet excitations, while remaining gapfull for the spin excitations (at least for $K\geq 10$, see Fig.~\ref{fig:spingap_vs_Klein}), forming an unusual spin liquid. This is what we investigate in the next section by looking at the low-energy singlet spectra.

\subsection{Spectrum of CF models}\label{sec:spectrum}

\begin{figure}
\includegraphics[width=\columnwidth]{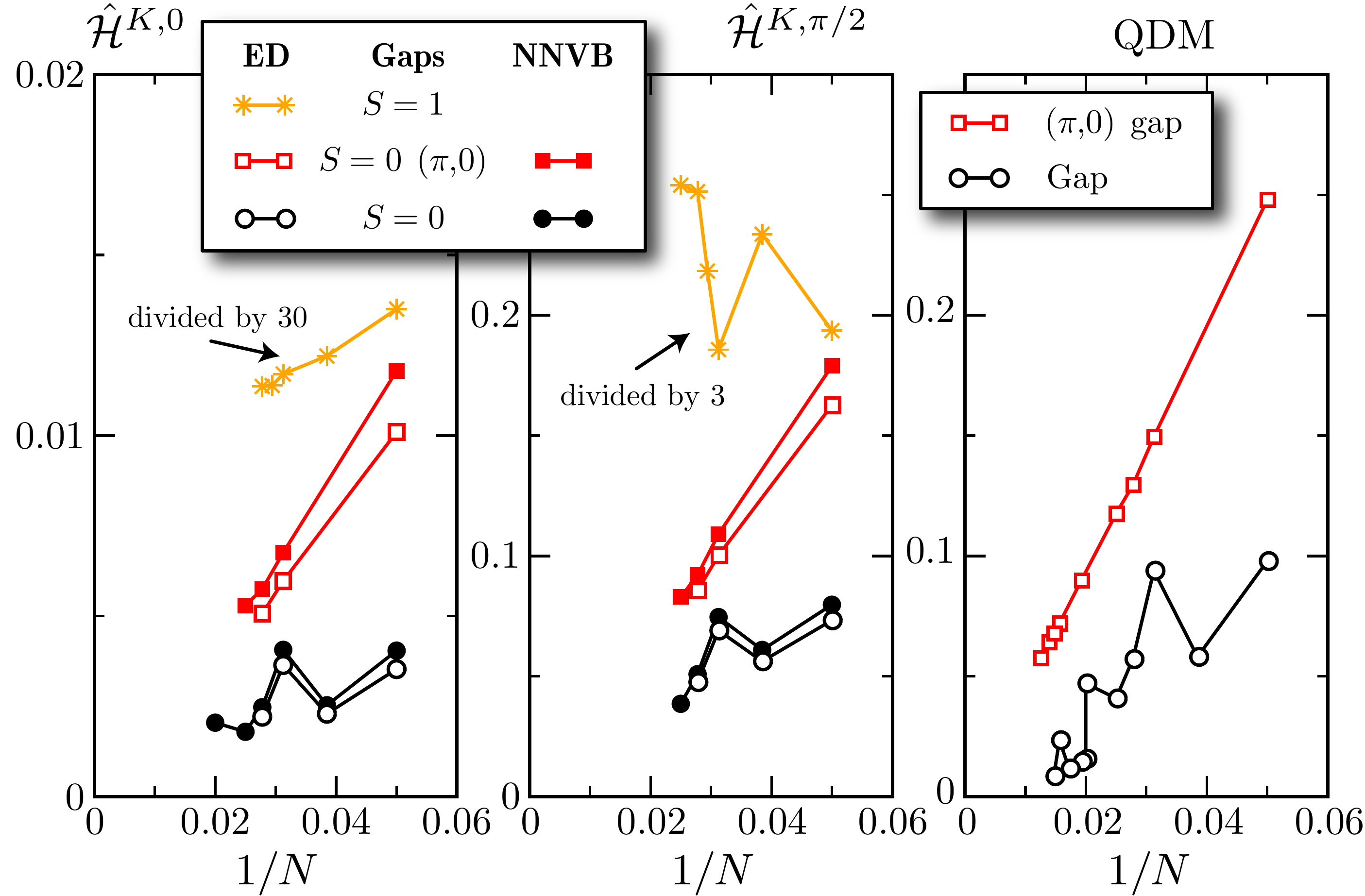}
\caption{(color online) Triplet and singlet gaps on various clusters for $\hat{\cal H}^{K,0}$ and $\hat{\cal H}^{K,\pi/2}$, using $K=10$. For comparison, we also plot the low-energy excitations of the QDM on the square lattice.}
\label{fig:Fig_pureCF}
\end{figure}

{\it Spin gap}. A shown in Fig.~\ref{fig:Fig_pureCF} (left and middle panels), for both $\hat{\cal H}^{K,0}$ and $\hat{\cal H}^{K,\pi/2}$ (with from now on, $K=10$), the triplet energy scale is well decoupled from low energy singlet excitations. For $\hat{\cal H}^{K,0}$, our data are compatible with the existence of a sizable spin gap in the thermodynamic limit. Its precise value is not an important property of the construction since, as shown in the previous subsection, it can be easily modified by an appropriate tuning of $K$. A similar behavior is observed for $\hat{\cal H}^{K,\pi/2}$ despite a less clear finite size scaling.

{\it Singlet excitations}. In order to characterize singlets excitations of  $\hat{\cal H}^{K,0}$ and $\hat{\cal H}^{K,\pi/2}$, we performed momentum resolved exact diagonalizations up to $N=36$ sites and NNVB diagonalizations up to $N=50$ sites. Before entering into the discussion of the results, let us make an important comment on the NNVB approach in the context of the two models studied here. The NNVB calculation, as it relies on a truncation of the singlet subspace, leads to a variational approximation of the GS energy in each symmetry sector. The singlet gap being a difference of two variational energies, its value cannot be strictly interpreted neither as an 
upper nor as a lower bound in general. Here, the situation is different as the zero-energy ground states are {\em exactly} captured by the NNVB approach. The singlet gap value obtained from this approach is therefore also variational, providing an {\em upper-bound} for the exact singlet gap.

The lowest singlet excitation, represented on Fig.~\ref{fig:Fig_pureCF} (left and middle panels), shows for both models a non-monotonous scaling as the system size increases. This can be ascribed to the fact that the first singlet excitations is not located at the same $\mathbf{k}$-point from one cluster size to another. For each size where ED and NNVB can be directly compared, both calculations leads to the same $\mathbf{k}$-point location of the gap and the data show that the NNVB approximation provides a very good upper bound of the singlet gap. Nevertheless, even by performing NNVB diagonalizations up to $N=50$ sites, the finite size scaling is still rather inconclusive although compatible with $0$ in the thermodynamic limit.

In contrast, the situation for the $(\pi,0)$ singlet gap is much clearer : both ED and NNVB data exhibit a linear scaling in $1/N$ which convincingly extrapolates to $0$ for infinite systems. $\hat{\cal H}^{K,0}$ and $\hat{\cal H}^{K,\pi/2}$ are thus gapless in the singlet sector. 

{\it QDM comparison}. The low energy singlet spectrum structure discussed above, is very reminiscent of the QDM spectrum at the RK point~\cite{Moessner2003} and in particular the presence of the pi0ns ({\it i.e.} $(\pi,0)$) low-lying branch. The right panel of Fig.~\ref{fig:Fig_pureCF} displays the finite-size scaling of the lowest gap and the $(\pi,0)$ gap for the QDM up to $N=80$ sites. This indeed confirms that, at a qualitative level, $\hat{\cal H}^{K,0}$, $\hat{\cal H}^{K,\pi/2}$ and the QDM at the RK point share similar low-energy excitations, where even the finite size scaling non-monotonicity is reproduced. In fact, because of the quadratic dispersion of the resonons at $(\pi,\pi)$, one expects a gapless spectrum at \emph{any momentum} ${\bf k}$.~\cite{Laeuchli2008}

{\it Comparison of energy scales in $\hat{\cal H}^{K,0}$ and $\hat{\cal H}^{K,\pi/2}$}. Even if the two models mimic the same dynamics for nearest neighbor SU(2) dimers (see section \ref{sec:model} and in particular Eq.~(\ref{eq:dyn1})-(\ref{eq:dyn2})), the two models remain different due to the presence of spin-$3/2$ projectors on the opposite corners of square plaquettes that involve non-trivial NNVB reconfigurations outside flippable plaquettes. Comparing $\hat{\cal H}^{K,0}$ and $\hat{\cal H}^{K,\pi/2}$ data reveals (see Fig.~\ref{fig:Fig_pureCF} left and middle panels) that the main difference is the energy scale at which singlet excitations occur, the gaps of $\hat{\cal H}^{K,0}$ being an order of magnitude smaller than those of $\hat{\cal H}^{K,\pi/2}$. This intriguing difference, that seems difficult to explain with simple arguments, will be addressed in the next part using a GQDM mapping. We will show that it can be ascribed to the presence of effective larger loops processes corrections to the QDM dynamics which are implicitly encoded in the spin projectors and are different in magnitude for the two models.

{\it Conclusion}. Our results indicate that $\hat{\cal H}^{K,0}$, $\hat{\cal H}^{K,\pi/2}$ and the QDM at the RK point have a rather similar low-energy structure. It is therefore tempting to extend the CF construction to build a family of SU(2) invariant spin-$1/2$ models which low-energy properties will mimic those of the QDM at an arbitrary value of $v/\vert t \vert$.

\section{Arbitrary $\theta$ models}\label{sec:mixingCF}

Having established the physical properties of the two models $\hat{\cal H}^{K,0}$, $\hat{\cal H}^{K,\pi/2}$ that were introduced by Cano and Fendley~\cite{CF}, we now move to arbitrary mixing angle $\theta$ realizations and show that they can host \emph{any} phase realized in their QDM counterparts. To motivate this relationship, we first derive an effective dimer model using the GQDM formalism. We then perform direct numerical simulations of $\hat{\cal H}^{K,\theta}$ for three different values of $\theta$.  As discussed in Sec.~\ref{sec:Klein}, we need for these simulations to consider a proper amplitude $K$ for the Klein term, and we again found that setting $K=10$ is sufficient (i) to ensure a finite spin gap for all $\theta$ that we will consider in this section (data not shown), and (ii) to be handled numerically without introducing a too large cost.

\subsection{GQDM approach}

{\it GQDM mapping}. In some recent works\cite{Ralko2009,Poilblanc2010,Schwandt2010,Albuquerque2011}, a general scheme allowing to perform analytically the projection of any SU(2) spin model on the non-orthogonal NNVB manifold was developped and applied to various cases. The effective models obtained are defined in the orthogonal basis of nearest neighbors hardcore dimers and only involve (i) potential (diagonal) terms sensitive to the possibility of flipping dimers along a plaquette of a given shape (ii) kinetic (off-diagonal) terms that actually perform the flip along a plaquette. This type of generalized QDM hamiltonian differs from the QDM Eq.~(\ref{eq:HQDM}) by the fact that flips are no longer limited to square plaquettes.

The shape of the considered plaquettes depends on the lattice and on the precision required for the projection. Since the amplitude of plaquette terms in the effective hamiltonian decreases exponentially with the number of dimers involved in the flipping process, it is generally sufficient to truncate the hamiltonian retaining only the smallest loops contributions. Recently, other schemes to derive GQDM from Heisenberg model have been proposed claiming to refine the procedure in the context of the kagome and square-kagome antiferromagnets~\cite{Ioannis2014,Ralko2015}. One has still however to perform a direct numerical comparison with the parent spin hamiltonian to understand whether its physical properties are well reproduced by the effective models. Here we will use the original GQDM scheme\cite{Schwandt2010} and make a precise connection between the spin hamiltonian $\hat{\cal H}^{K,\theta}$ Eq.~(\ref{eq:generalmodel}) and the QDM Eq.~(\ref{eq:HQDM}) that is fully confirmed by exact diagonalizations.

On the square lattice, the smallest plaquette terms are~:
\begin{align}
\label{eq:plaq4kin}
\sq{1_1} &=  \state[sq1]{0.6cm}\;  \state[qs2]{0.6cm} + \; \state[sq2]{0.6cm} \; \state[qs1]{0.6cm}, \\
\label{eq:plaq4pot}
\sq{1_2} &= \state[sq1]{0.6cm} \; \state[qs1]{0.6cm} + \; \state[sq2]{0.6cm} \; \state[qs2]{0.6cm}.
\end{align}

Here, we will also consider next-leading plaquette terms~:
\begin{align}
\label{eq:plaq6akin}
\sq{2_2} &=  \state[rec1]{0.91cm}\;  \state[cer2]{0.91cm} + \; \state[rec2]{0.91cm} \; \state[cer1]{0.91cm}, \\
\label{eq:plaq6apot}
\sq{2_6} &= \state[rec1]{0.91cm}\;  \state[cer1]{0.91cm} + \; \state[rec2]{0.91cm} \; \state[cer2]{0.91cm}, \\
\label{eq:plaq6bkin}
\sq{2_4} &=  \state[rec1]{0.91cm}\;  \state[cer3]{0.91cm} + \; \state[rec3]{0.91cm} \; \state[cer1]{0.91cm}, \\
\label{eq:plaq6bpot}
\sq{2_7} &= \state[rec1]{0.91cm}\;  \state[cer1]{0.91cm} + \; \state[rec3]{0.91cm} \; \state[cer3]{0.91cm}.
\end{align}

\begin{table*}
  \centering
\begin{tabular}{lcrp{0.5cm}crp{0.5cm}crp{0.5cm}crp{0.5cm}cr}
\hline\hline
&\multicolumn{2}{c}{$\hat{\cal{H}}^0$} & & \multicolumn{2}{c}{$\hat{\cal{H}}^1$} & & \multicolumn{2}{c}{$\text{\bf A}$}& & \multicolumn{2}{c}{$\text{\bf B}$}& & \multicolumn{2}{c}{$\text{\bf C}$}\\
&\multicolumn{2}{c}{$\theta=0$} & & \multicolumn{2}{c}{$\theta=\pi/2$} & & \multicolumn{2}{c}{$\theta_A=-\arccos\left(-\frac{3}{\sqrt{10}}\right)$}& & \multicolumn{2}{c}{$\theta_B=-\arccos\left(\frac{3}{\sqrt{10}}\right)$}& & \multicolumn{2}{c}{$\theta_C=\arccos\left(\frac{1}{\sqrt{10}}\right)$}\\[3pt]
\hline
$t_4$  								&& $0.25$        &&& $-0.25$      &&& $-0.125$			&&& $0.25$				&&& $-0.125$			\\
$v_4$  								&& $0.25$        &&& $0.25$       &&& $-0.25$				&&& $0.125$				&&& $0.25$				\\
$t_6$  								&& $-0.0885651$  &&& $0.202914$   &&& $0.015695$		&&& $-0.117152$		&&& $0.130044$		\\
$v_6$  								&& $-0.031203$   &&& $-0.0308274$ &&& $0.0311091$		&&& $-0.0156954$	&&& $-0.0309213$	\\
$t'_4$								&& $0.063808$    &&& $0.0742716$  &&& $-0.0664239$	&&& $0.0292881$		&&& $0.0716557$		\\
$v'_4$ 								&& $0.0364349$   &&& $0.0779137$  &&& $-0.0468046$	&&& $0.00784772$	&&& $0.067544$		\\
$v_4/\vert t_4 \vert$ &&   $1$         &&& $1$					&&& $-2.$					&&& $0.5$					&&& $2.$					\\
\hline\hline
\end{tabular}
  \caption{GQDM amplitudes (see text) for several specific values of the mixing angle $\theta$.}
  \label{tab:GQDMCoeff}
\end{table*}

The effective GQDM hamiltonian is defined as ${\cal O}^{-1/2} {\cal H}^{K,\theta} {\cal O}^{-1/2}$, where ${\cal O}$ and ${\cal H}$ are the overlap and hamiltonian operators : 
\begin{align}
\label{eq:o}
{\cal O} &=  \alpha^2 \sq{1_1} + \alpha^4 \sq{2_2} + {\text{ larger loops }}, \\
\label{eq:h}
{\cal H}^{K,\theta} &= h_{\,\state[D_1_1]{0.15cm}} \alpha^2 \sq{1_1}+h_{\,\state[D_1_2]{0.15cm}} \sq{1_2} + h_{\,\state[D_2_2]{0.3cm}} \alpha^4 \sq{2_2} \\
&+ h_{\,\state[D_2_6]{0.3cm}}  \sq{2_6} + h_{\,\state[D_2_4]{0.3cm}}  \alpha^2 \sq{2_4}+ h_{\,\state[D_2_7]{0.3cm}}  \sq{2_7} \nonumber\\
&+ {\text{ larger loops }} \nonumber,
\end{align}
with $\alpha^2=1/2$ (this value originating from the non orthogonality of the NNVB basis and the overlap rule), and the values of $h_{\state[D_1_1]{0.15cm}}, \ldots,  h_{\state[D_2_7]{0.3cm}}$ are determined by $\hat{\cal H}^{K,\theta}$. It is important to note that in the GQDM scheme\cite{Schwandt2010}, the above expressions are not series in $\alpha^2$ but expansions on GQDM operator basis. In other words, no assumption is made on $\alpha$.

The final effective Hamiltonian for ${\cal H}^{K,\theta}$ takes the form
\begin{align}
\label{eq:heff}
{\cal H}^{\text{eff},\theta} &= -t_4(\theta)  \sq{1_1}+v_4(\theta) \sq{1_2} - t_6(\theta) \sq{2_2} \\
&+ v_6(\theta) \sq{2_6} - t'_4(\theta) \sq{2_4} + v'_4(\theta) \sq{2_7} \nonumber \\
&+ {\text{ larger loops }} \nonumber
\end{align}

where the analytical expressions for parameters (which all depend on the mixing angle $\theta$) have been reported to the Appendix~\ref{appendix2} (see Eqs.~\ref{eq:t4}-\ref{eq:vp6}) for conciseness. A few specific values are discussed in the following (see Table~\ref{tab:GQDMCoeff}).

{\it Some remarks on $\hat{\cal H}^{K,0}$ and $\hat{\cal H}^{K,\pi/2}$}. Interestingly enough, the two models introduced by Cano and Fendley\cite{CF}, which correspond to $\theta=0$ and $\pi/2$, exactly map to the same RK hamiltonian $\vert t_4 \vert = v_4 =1/4$ if truncated to the first relevant terms $t_4$ and $v_4$. Pushing the mapping further reveals differences between $\hat{\cal H}^{K,0}$ and $\hat{\cal H}^{K,\pi/2}$, as they for instance introduce different dynamics (such as the $t_6$ term) in the NNVB manifold.

\begin{figure}
  \begin{center}
    \includegraphics*[width=\linewidth]{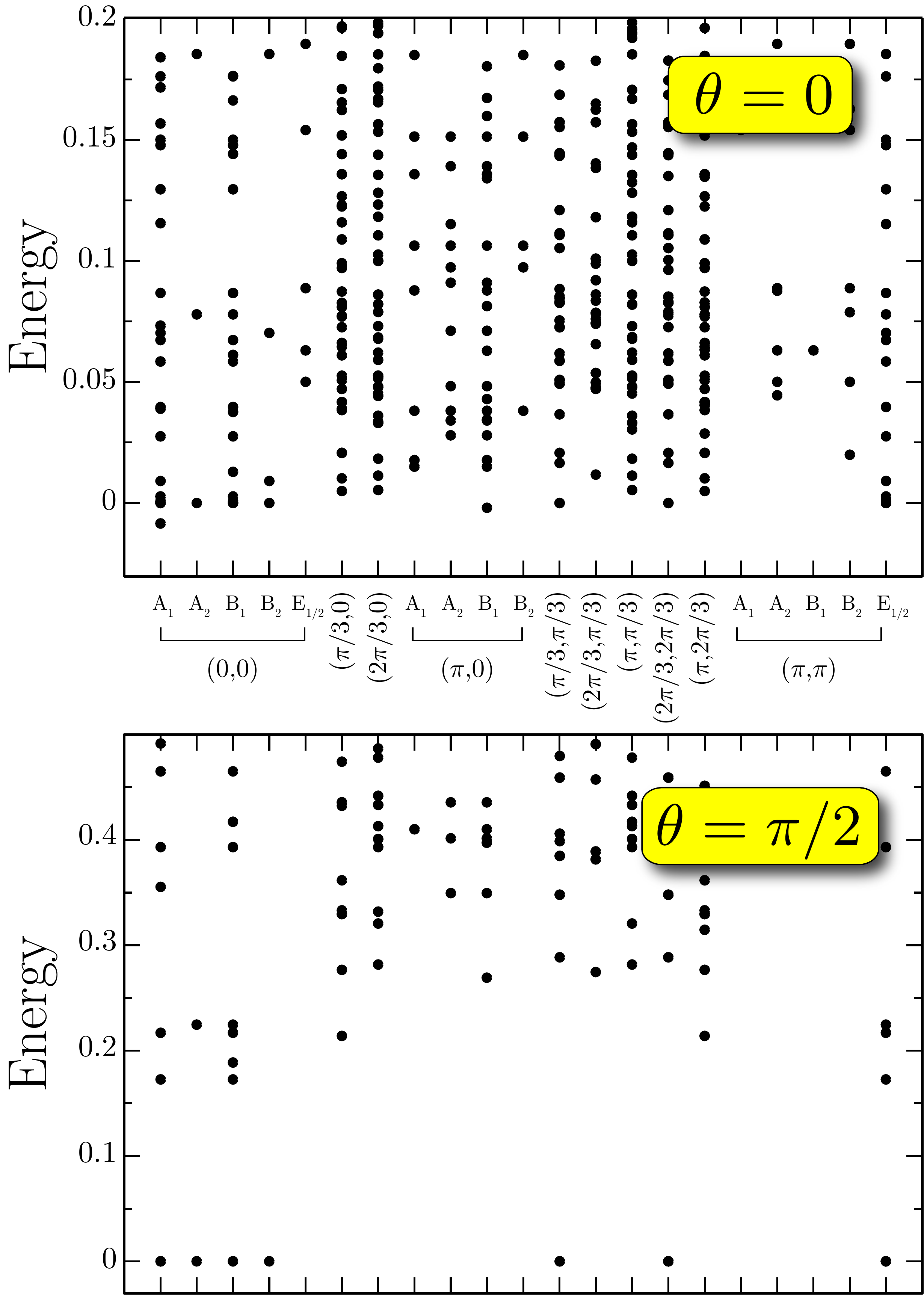}
   \end{center}
   \caption{(color online) Exact spectra of the GQDM models at $\theta=0$ (top panel) and $\theta=\pi/2$ (bottom panel) for a 36-site cluster. States are labelled according to their momentum (one data set per nonequivalent ${\bf k}$ point). For ${\bf k}=(0,0)$ (respectively $(\pi,\pi)$), we also add a label corresponding to the irreducible representation of C$_{4v}$ (respectively C$_{2v}$) point group symmetry.}
\label{fig:GQDM_CF0_CF1_exact}
\end{figure}

As to prove the ability of the GQDM mapping to capture the properties of $\hat{\cal H}^{K,\theta}$, we first investigate whether the corrections $t_6(\theta)$, $v_6(\theta)$, $t'_4(\theta)$, $v'_4(\theta)$ to the QDM can explain the difference in the low-lying singlet excitations energy scale between $\hat{\cal H}^{K,0}$ and $\hat{\cal H}^{K,\pi/2}$ observed in section \ref{sec:spectrum}. On Fig.~\ref{fig:GQDM_CF0_CF1_exact} the ED spectra of GQDM models corresponding to  $\hat{\cal H}^{K,0}$ and $\hat{\cal H}^{K,\pi/2}$ are represented for a 36-site cluster in each symmetry sector. The data show indeed a clear difference in energy scales of the lowest excitations, very similar to the results obtained with ED and NNVB calculations on the original spin models (see Fig.~\ref{fig:Fig_pureCF}).

{\it Exploring $v/\vert t \vert$ by varying $\theta$}. Using the expressions of $v_4 (\theta)$ and $t_4 (\theta)$ reported in Eqs.~(\ref{eq:t4}) and (\ref{eq:v4}) of Appendix~\ref{appendix2}, we represent on Fig.~\ref{fig:GQDMCoeff} the variation of the ratio $v_4/|t_4|$ as a function of $\theta$. As can be seen from this figure, any value of the ratio $v_4/\vert t_4 \vert$  can be reached by an appropriate tuning of $\theta$. The value of $v_4/|t_4|$, together with the knowledge of the QDM phase diagram, serves as a guideline to understand which  phases are accessible for the ground-state of $\hat{\cal H}^{K,\theta}$ for a given $\theta$. The phase diagram of the square lattice QDM exhibits VBC states~\cite{Sachdev1989,Leung1996,Syljuasen2006} such as columnar for $v/|t|<0$ and staggered for $v/|t|>1$. In the intermediate region, a plaquette phase has been suggested to exist~\cite{Sachdev1989,Leung1996,Syljuasen2006}, or a mixed columnar-plaquette~\cite{Ralko2008a}, but its presence has recently been questioned and the columnar phase seems to extend up to the RK point~\cite{Banerjee2014,Schwandt2015}. 

Of course larger loops processes, in particular $t_6(\theta)$, must also be taken into account and can somehow modify this simple picture. A careful numerical study of $\hat{\cal H}^{K,\theta}$ should therefore be pursued, which will be done using both exact and NNVB diagonalization, together with an exact diagonalization study of the GQDM counterpart. In the following we will concentrate on three particular values of $\theta$ denoted $\theta_A$, $\theta_B$ and $\theta_C$ (see Table \ref{tab:GQDMCoeff} and Fig.~\ref{fig:GQDMCoeff}), for which we can expect respectively  columnar VBC, putative mixed plaquette-columnar VBC and staggered VBC phases from the $v_4/\vert t_4 \vert$ QDM proxy.

\begin{figure}
  \begin{center}
    \includegraphics*[width=\linewidth]{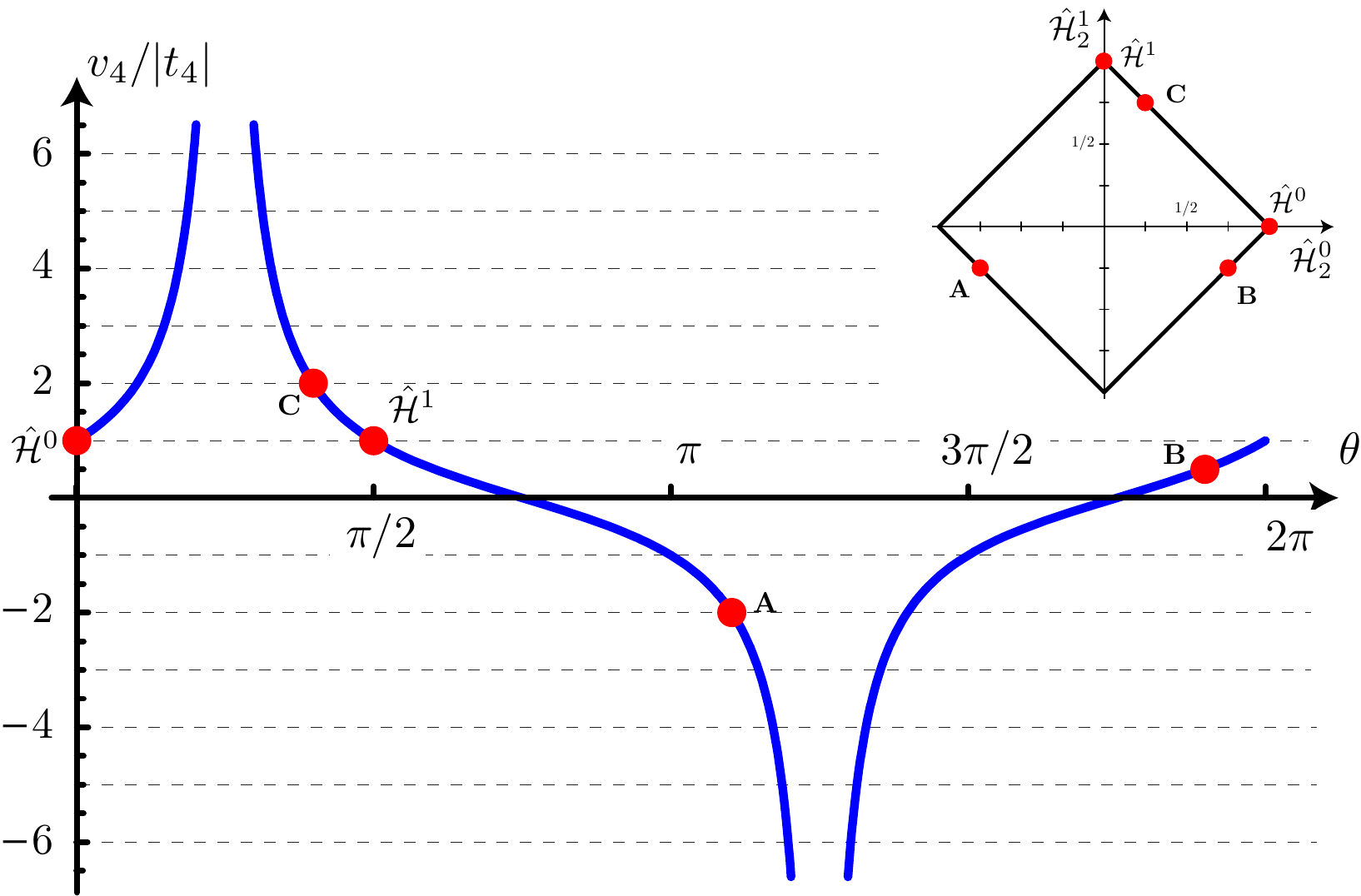}
   \end{center}
   \caption{(color online) Variation of $v_4/\vert t_4 \vert$ as a function of $\theta$. Five specific values of $\theta$ are displayed as red dots corresponding to $\hat{\cal H}^{K,0}$, $\hat{\cal H}^{K,\pi/2}$ and 3 other values of $v_4/\vert t_4 \vert$ studied in the main text.}
  \label{fig:GQDMCoeff}
\end{figure}


\subsection{Low-energy excitations and dimer correlations}
In order to characterize possible symmetry breaking, it is useful to
investigate the low-energy levels quantum numbers on a finite
cluster. The VBC phases we expect for models A, B and C will have clear signature in the low-energy spectra as singlet excitations with definite quantum numbers that should collapse onto the ground-state in the thermodynamic limit.~\cite{Ralko2008a} 
Moreover, dimer ordering can also be checked directly by computing (connected) dimer-dimer correlations on the finite-size ground-state.
We describe in detail below the low-energy physics of model A, B and C, based on the collection of our numerical data on a $N=36$ square cluster presented in Fig.~\ref{fig:phase_diagram}.

\subsubsection{Model $A$ using $\theta_A$}
\begin{figure*}[!ht]
  \begin{center}
    \includegraphics*[width=0.99\linewidth]{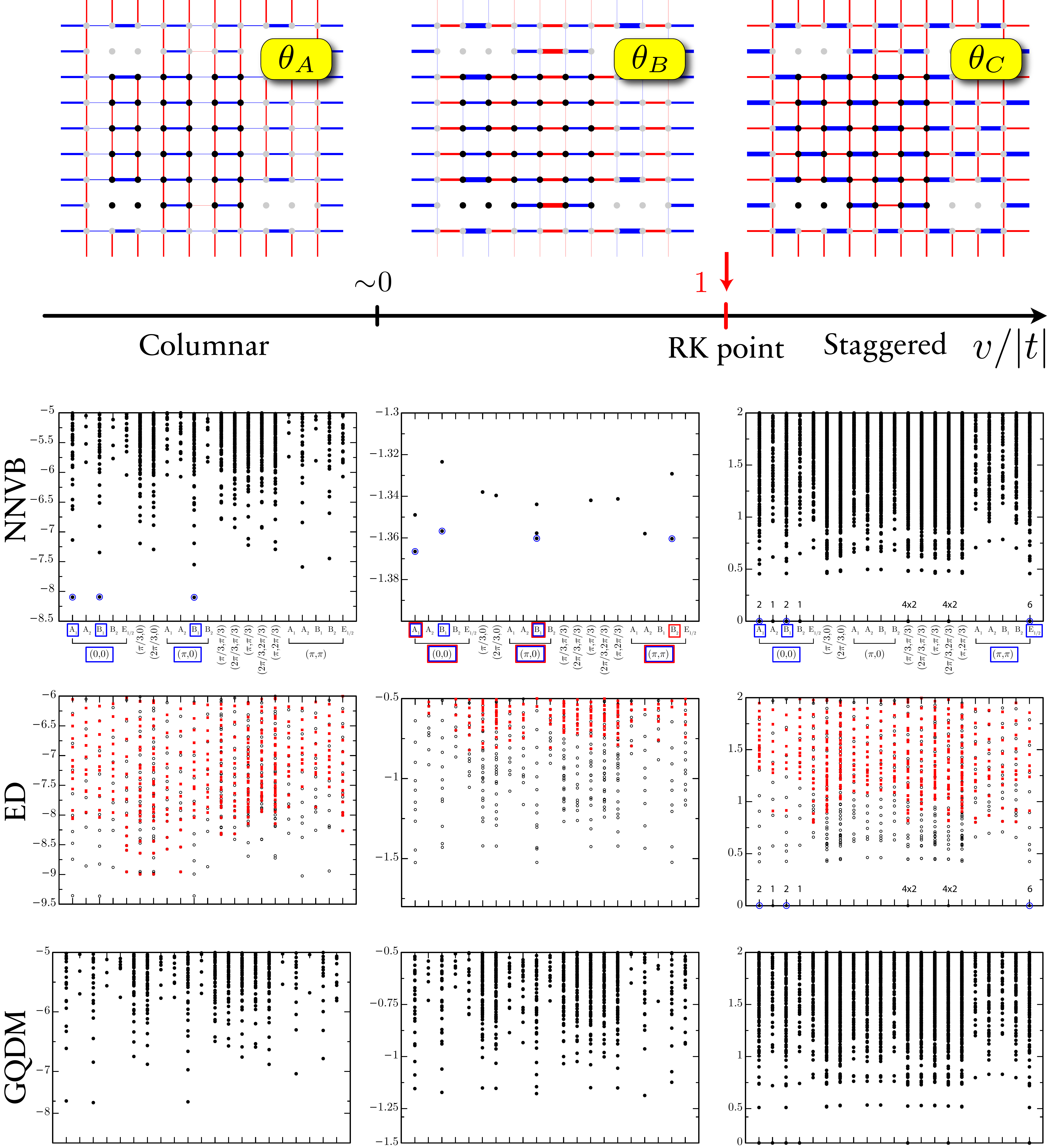}
   \end{center}
   \caption{(color online) Phase diagram of the QDM (second top panel), dimer correlations (first top panels, for clarity only NNVB data on $N=36$ cluster) and low-energy spectra of ${\cal H}^{K,\theta}$ from exact (fourth panels from top) and NNVB diagonalization (third panels from top), and of the related GQDM (fifth panels from top). For dimer correlations, blue (red) bonds denote positive (negative) correlations, while the width of each bond is proportional to $C_{ijkl}$ (see main text for definition). The energy spectra are represented as a function of the wave-vector and point-group symmetry groups of the $N=36$ square lattice cluster (see Fig.~\ref{fig:GQDM_CF0_CF1_exact} for labels). The squares surrounding the wave-vector and point group labels highlight the sectors for which low-lying states are expected in order to form a columnar state (blue squares of the left panel), staggered state (blue squares on the right panels) and a columnar or plaquette state (respectively blue and red squares on the middle panel). The corresponding energy levels are pointed out with a circle. ED data are also labeled differently for $S_z=0$ and $S_z=1$ (black circle and red square respectively).
}
  \label{fig:phase_diagram}
\end{figure*}

According to table~\ref{tab:GQDMCoeff}, such a combination would map onto a related QDM model with $v_4/|t_4|=-2$ which is known to be deep in the columnar phase.~\cite{RK} Indeed, Fig.~\ref{fig:phase_diagram}(a) shows the low-energy spectra obtained on a $N=36$ cluster by ED and NNVB simulations for the spin model as well as for the GQDM for the point $\theta_A$. In all spectra, one can easily identify four almost-degenerate ground-states with quantum numbers corresponding to columnar order (i.e. $(0,0)$ $A_1$, $(0,0)$ $B_1$ and two-fold degenerate $(\pi,0)$ $B_1$) and a sizable gap above them. It is also remarkable that a large part of the low-lying singlet excitations structure is common between the ED and NNVB computations and is also reproduced by the GQDM. 

The top panel of Fig.~\ref{fig:phase_diagram}(a) represents graphically the dimer correlations $C_{ijkl}=\langle ({\bf S}_i \cdot {\bf S}_j)  ({\bf S}_k \cdot {\bf S}_l) \rangle -  \langle {\bf S}_i \cdot {\bf S}_j \rangle \langle {\bf S}_k \cdot {\bf S}_l \rangle $, taken with respect to a reference bond with NN sites $(i,j)$ denoted with black dots, and where the position of the NN sites $(k,l)$ is varied. This calculation is done using the ground-state in the NNVB basis. One clearly observes a columnar pattern, in agreement with the analysis above.

\clearpage
\subsubsection{Model $C$ using $\theta_C$}
One expects to engineer this way a spin model mimicking the QDM with $v_4/|t_4|=2$ which is known to 
possess an exact staggered ground-state. Let us mention that other non flippable zero-energy states can be constructed so that the actual degeneracy of the QDM is extensive: for instance, we find 28 zero-energy states on $N=36$ cluster for the QDM model with these couplings. Those include not only the 4 states $(0,0) A_1$, $(0,0) B_1$, $(\pi,\pi) E_1$ and $(\pi,\pi) E_2$ forming the staggered "pure" states but a total of $4(2^{L/2}-1)$ states (with $N=L \times L$) all belonging to the the diagonals of the Brillouin zone (see Appendix~\ref{appendix1}). The overall degeneracy of $28$ is decomposed in the following way: 6 states with momentum $(0,0)$, 8 in $(\pm \pi/3, \pm \pi/3)$, 8 in $(\pm 2\pi/3,2 \pm \pi/3)$ and 6 in $(\pi,\pi)$.

In Fig.~\ref{fig:phase_diagram}(c), one can readily observe that the low-energy spectra   
of the spin-$1/2$ model (both with ED and in the NNVB basis) as well as of the GQDM perfectly display this exact same degeneracy. Again, the first non-zero singlet excitations are also identical between ED and NNVB, and well reproduced by the GQDM.

With such a ground-state degeneracy, the correlations depend on the chosen ground-state. 
Taking the fully symmetric ground-state (i.e. with momentum $(0,0)$ and fully symmetric with respect to the C$_{4v}$ point group), the existence of a staggered phase is confirmed by a direct computation of the dimer correlations in Fig.~\ref{fig:phase_diagram}(c).

\subsubsection{Model $B$ using $\theta_B$}
As a last example, we studied ${\cal H}^{K,\theta_B}$ (see table~\ref{tab:GQDMCoeff}) which is expected to mimic QDM at $v_4/|t_4|= 1/2$, i.e. in a region where the columnar order is not fully established yet even using large-scale quantum Monte-Carlo simulations.~\cite{Banerjee2014,Schwandt2015}

Fig.~\ref{fig:phase_diagram}(b) presents the low-energy spectra where we observe in ED and NNVB computations and GQDM spectra, that both set of states expected to construct columnar or plaquette states in the thermodynamic limit are present at low-energy. Note that other states in different sectors not expected to contribute to columnar or plaquette ordering (such as {\it e.g.} ${\bf k}=(\pi/3,0)$ or ${\bf k}=(2\pi/3,0)$) also have low energy. The competition  on finite clusters between these different states make for a slightly not as good agreement between the three types of spectra for this model with $\theta_B$, as compared to $\theta_A$ and $\theta_C$ where the low-energy picture is clearer.

A similar observation of low-energy states in competition (between the columnar and plaquette states) can be made on the related QDM model and had led to the conclusion that it could be a signature of a mixed phase.~\cite{Ralko2008a} However, in the case of the QDM, there is no clear gap above these eight states and refined ED and QMC studies have found no evidence for such mixed phase~\cite{Banerjee2014,Schwandt2015}: it is instead indicative of a U(1) regime where no crystalline order is frozen yet (on the available system sizes). Presumably, the system sizes that can be reached are too small to give any reliable information on the thermodynamic limit because of the existing diverging length close to RK point.~\cite{Banerjee2014,Schwandt2015}

While in principle, the supplementary terms ($t_6,v_6$ etc) induced by the non-trivial singlet dynamics in ${\cal H}^{K,\theta}$ could favor one particular type of ordering, we cannot make either any definitive statement on the behavior in the thermodynamic limit for the point $\theta_B$, based on our finite cluster simulations. This is confirmed by computations of correlations in Fig.~\ref{fig:phase_diagram}(b), which indicate also a kind of mixed phase with both columnar and plaquette patterns (see Ref.~\onlinecite{Mambrini2006} for a related discussion).

\section*{Conclusion}\label{sec:conc}

Using numerically exact simulations as well as an analytical mapping to a generalized quantum dimer model, we have studied in detail the low-energy properties of fully-SU(2) symmetric spin-$1/2$ models introduced by Cano and Fendley on the square lattice~\cite{CF}. We confirm that the original models in Ref.~\onlinecite{CF} indeed admit equal-amplitude superpositions of nearest-neighbor valence bond states, which is what they were designed for. Besides the corresponding ground-state degeneracy, our calculations show that the low-energy physics is also in correspondence with the one encountered in the quantum dimer model. More precisely, the CF models admit gapless spinless excitations while having a spin gap in the thermodynamic limit (for sufficiently large Klein term), thereby forming an unusual type of spin liquid (the ground-state displays power-law dimer correlations).

Mixing the two models, provides {\it a SU(2) spin-1/2 hamiltonian analog to QDM} : we find that we can reproduce quite trustfully the phases encountered in the quantum dimer model, namely columnar, staggered (and apparent plaquette or mixed) valence bond crystals in these spin-$1/2$ models. This strongly suggests that the CF models are multicritical points similar to the Rokhsar-Kivelson point of the quantum dimer model, and that the unusual spin liquid encountered there is not generic.

Our work can possibly be extended in several different directions. First, it would be interesting to see whether techniques specific to the RK point of the QDM, could also be used for CF models: for instance, we could try to construct variational states using a single mode approximation to understand more precisely the spinless excitations that become gapless in the thermodynamic limit~\cite{Moessner2003}. Also, it would be interesting to see whether exact dynamics could be performed, at least in the NNVB sector, using the techniques introduced in Ref.~\onlinecite{Henley97}.

Secondly and away from the RK point, it would be interesting to perturb the columnar crystal (as e.g. found for $\theta_A$) with an antiferromagnetic Heisenberg exchange between NN spins, which is now possible since we directly have access to spins. The possibility of a  deconfined quantum critical point~\cite{Senthil04} between the N\'eel and VBC phases can be explicitly studied, albeit on the relatively small sizes available to exact diagonalization.

Finally, a construction similar to the one introduced by CF for the square lattice, can be performed on non-bipartite lattices, where the QDM exhibits $Z_2$ liquid phases close and at the RK point~\cite{Moessner2001a,Misguich2002}. For instance, we are
currently checking  whether the ground-state of a CF-like model on the kagome lattice is adiabatically connected to the one of the Heisenberg model, which has been recently argued to host a $Z_2$ spin
liquid.~\cite{Yan11,Jiang2012,Depenbrock2012} 

\section*{Acknowledgements}
We acknowledge useful comments from Jennifer Cano, Paul Fendley and Didier Poilblanc. S. C. would like to thank Andreas L\"auchli for help in simulating QDM models and related collaboration.
 This work has been supported by 
the French ANR program ANR-08-JCJC-0056-01, Institut Universitaire de 
France (SC) and Indo-French Centre for the Promotion of Advanced Research (IFCPAR/CEFIPRA) under Project 4504-1 (FA). Numerical simulations have been performed using resources from GENCI--CCRT, GENCI--IDRIS (grant x2015050225) and CALMIP.

\appendix
\section{Ground-state degeneracies of the RK point on the square lattice}\label{appendix1}

\begin{figure}
\includegraphics[width=\columnwidth]{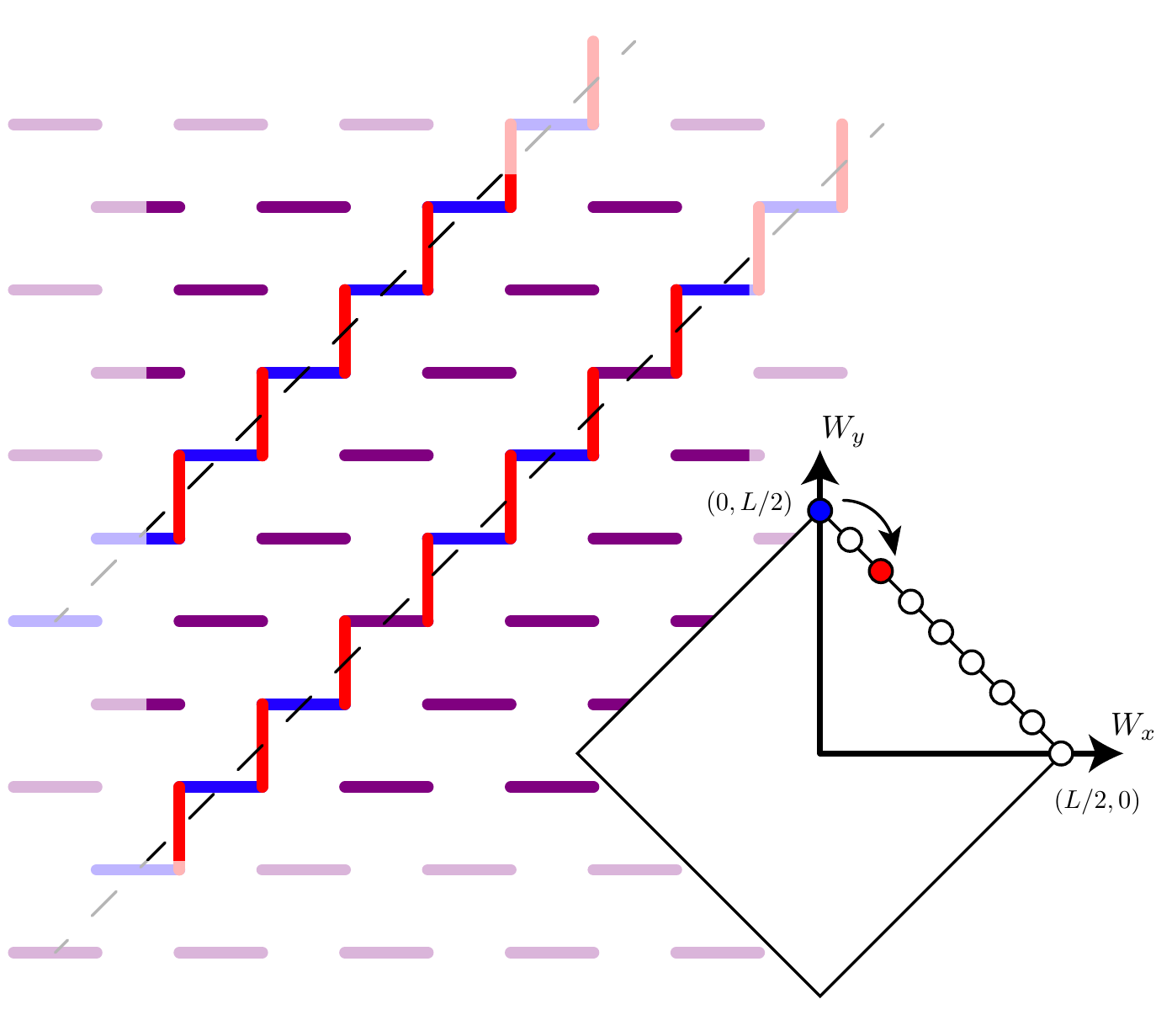}
\caption{(color online) In blue, one staggered dimer configuration with winding $(0,L/2)$. In red, by flipping dimers along two diagonal lines, one can generate another non-flippable configuration in the sector $(2,L/2-2)$. Inset shows the allowed winding sectors inside the $0\leq |W_x|+|W_y|\leq L/2$ square, as well as the sectors corresponding to these two configurations.}
\label{fig:Non_Flippable}
\end{figure}

\begin{figure}
\includegraphics[width=0.7\columnwidth]{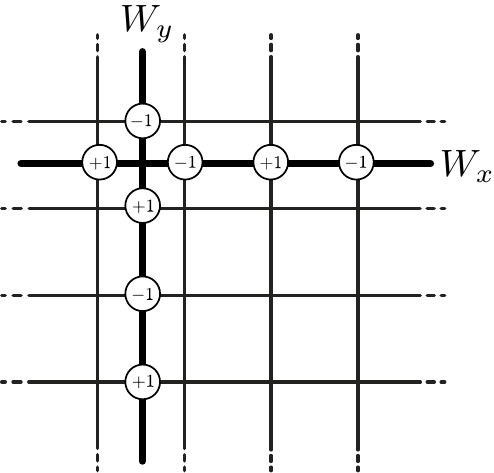}
\caption{(color online) Example of the sign convention used to determine the winding numbers $(W_x,W_y)$ of a dimer configuration (see text).}
\label{fig:Winding_convention}
\end{figure}

Let us consider the QDM at the RK point on the square lattice. It is well known that the equal superposition of \emph{all} dimer configurations is a ground-state of the hamiltonian. However, due to the existence of several ergodic sectors, there are in fact many ground-states and the purpose of this appendix is to compute this degeneracy explicitly, something was never presented to the best of our knowledge.

For a given $L\times L$ square lattice with periodic boundary conditions, dimer configurations can be labelled using their winding numbers $(W_x,W_y)$, $0\leq |W_x|+|W_y|\leq L/2$, see Fig.~\ref{fig:Non_Flippable}. $W_x$  (respectively $W_y$) can be easily computed by summing, on each link (with an alternating sign) cutting a line along $x$ (respectively $y$) direction, the contribution $\pm 1$ if the bond is occupied or 0 if it is empty (see Fig.~\ref{fig:Winding_convention} for the sign convention). It is straightforward to see that these labels are invariant with respect to any \emph{local} rearrangement of the dimer, i.e. they are conserved quantities for the Hamiltonian. So, there are at least $(L^2+2L+2)/2$ different sectors. 

Now, in each \emph{ergodic} sector, it is clear that the uniform state (equal weight superposition of all states in this sector) is a 
ground-state, that can be shown to be unique using Perron-Frobenius theorem. We have observed numerically on finite clusters that all sectors verifying $
0\leq |W_x|+|W_y|< L/2$ are indeed ergodic, and thus host one single ground-state each. However, the sectors corresponding to the boundary ($|W_x|+|W_y|=L/2$) behave differently. As shown in Fig.~\ref{fig:Non_Flippable}, starting from a non-flippable staggered dimer configuration (for instance, the unique state in the sector $(0,L/2)$), one can flip dimers along an arbitrary number of diagonal lines to generate other non-flippable configurations (i.e. zero energy states and thus ground-states) in other sectors along the boundary. Using simple combinatorics, one can construct ${{L/2}\choose{p}}=(L/2)\,!/(p\,!(L/2-p)\,!)$ configurations in the sector $(p,L/2-p)$ for instance. 

As a result, we can construct several ground-states and we predict a total degeneracy of $(L^2-2L-6)/2+2^{L/2+2}$. We have numerically checked that these are indeed the only ground-states (for instance 17 for $L=4$, 41 for $L=6$ and so on). In particular, the degeneracy is \emph{exponential} with the linear size (while the number of winding sectors is only polynomial). This may explain  why perturbing the RK point can lead to a variety of phenomena.~\cite{Fradkin04}

This counting also provides the number of \emph{non-flippable states}, that belong to the winding sectors along the edge, equal to $4(2^{L/2}-1)$. As a final comment, since we cannot define winding sectors for a spin model, we can nevertheless use information about the momentum and/or point group symmetry of these states: by inspection, all these ground-states have momenta of the form $(\pm k,\pm k)$, i.e. along the diagonals of the Brillouin zone. 

\section{Derivation of the GQDM}\label{appendix2}
The GQDM hamiltonian is derived from the spin model using the techniques described in Ref.~\onlinecite{Schwandt2010}. It allows us to obtain the values of the GQDM coefficients including all orders in $\alpha^2$. For conciseness we only give in this appendix the key ingredients for the derivation of the GQDM based on hamiltonian Eq.~(\ref{eq:generalmodel}) and refer the reader to Ref.~\onlinecite{Schwandt2010} for extensive details on the technique. 

The first step relies in the direct evaluation of matrix elements of hamiltonian Eq.~(\ref{eq:generalmodel}) : 
\begin{align}
\label{eq:barevalues}
h_{\,\state[D_1_1]{0.15cm}} &=   \frac{1}{4n(\theta)} (-\cos \theta  + 3 \sin \theta ) \\
h_{\,\state[D_1_2]{0.15cm}} &=  \frac{1}{8n(\theta)} (\cos \theta  + 3 \sin \theta) \\
h_{\,\state[D_2_2]{0.3cm}} &=  h_{\,\state[D_2_6]{0.3cm}} = h_{\,\state[D_2_4]{0.3cm}} =h_{\,\state[D_2_7]{0.3cm}} =0 
\end{align}
with $n(\theta) = \vert \cos \theta \vert + \vert \sin \theta \vert$

After resummation of all contributions in $\alpha^2$, the coefficients of the model are:

\begin{align}
\label{eq:t4a}
t_4 &= \frac{1}{n(\theta)} \frac{3 \alpha ^2}{8 \left(1-\alpha ^4\right)}  (\cos \theta-\sin \theta)  \\
\label{eq:v4a}
v_4 & = \frac{1}{n(\theta)} \frac{1}{8 \left(1-\alpha ^4\right)}  (  (1+2 \alpha ^4) \cos \theta  + 3 (1-2 \alpha ^4)\sin \theta  )
\end{align}
\begin{multline}
\label{eq:t6a}
t_6=  \frac{1}{n(\theta)} \frac{1}{8 \left(8-7 \alpha ^4-\alpha ^8\right)} \times \\
\left (  \left ( 2 \left(\sqrt{1-\alpha ^4}-1\right)-\alpha ^4 \left(4 \sqrt{1-\alpha ^4}+13\right) \right ) \cos \theta  \right . \\
\left . + \left ( 3 \left(4 \sqrt{1-\alpha ^4}+11\right) \alpha ^4+6 \left(\sqrt{1-\alpha ^4}-1\right) \right ) \sin \theta   \right )
\end{multline}
\begin{multline}
\label{eq:v6a}
v_6 =  \frac{1}{n(\theta)} \frac{1}{8 \left(8-7 \alpha ^4-\alpha ^8\right)} \times \\
\left (   \left ( -5 \alpha ^8+2 \left(4 \sqrt{1-\alpha ^4}-7\right) \alpha ^4-4 \sqrt{1-\alpha ^4}+4 \right ) \cos \theta  \right . \\
\left . + \left ( 3 \left(3 \alpha ^8+\left(2-8 \sqrt{1-\alpha ^4}\right) \alpha ^4-4 \sqrt{1-\alpha ^4}+4\right) \right ) \sin \theta   \right )
\end{multline}
\begin{multline}
\label{eq:tp4a}
t'_4 =  \frac{1}{n(\theta)} \frac{1}{8 \left(8-7 \alpha ^4-\alpha ^8\right)} \times \\
\left (   \left ( \alpha ^2 \left(\left(2 \sqrt{1-\alpha ^4}+7\right) \alpha ^4-\sqrt{1-\alpha ^4}+5\right) \right ) \cos \theta  \right . \\
\left . + \left (  -3 \alpha ^2 \left(\left(2 \sqrt{1-\alpha ^4}+5\right) \alpha ^4+\sqrt{1-\alpha ^4}-5\right) \right ) \sin \theta   \right )
\end{multline}
\begin{multline}
\label{eq:vp6a}
v'_6 =  \frac{1}{n(\theta)} \frac{1}{8 \left(8-7 \alpha ^4-\alpha ^8\right)} \times \\
\left (   \left ( 3 \alpha ^8+\left(11-4 \sqrt{1-\alpha ^4}\right) \alpha ^4+2 \left(\sqrt{1-\alpha ^4}-1\right) \right ) \cos \theta  \right . \\
\left . + \left (  -3 \alpha ^8+3 \left(4 \sqrt{1-\alpha ^4}+3\right) \alpha ^4+6 \left(\sqrt{1-\alpha ^4}-1\right) \right ) \sin \theta   \right )
\end{multline}

Using the fact $\alpha^2=1/2$, these coefficients take the more compact form:
\begin{align}
\label{eq:t4}
t_4 &= \frac{1}{4n(\theta)} (\cos \theta-\sin \theta)\\
\label{eq:v4}
v_4 & = \frac{1}{4n(\theta)} (\cos \theta+\sin \theta)\\
\label{eq:t6}
t_6 &=  \frac{1}{198n(\theta)} \left(\left(2 \sqrt{3}-21\right) \cos \theta +9 \left(1+2 \sqrt{3}\right) \sin \theta \right)\\
\label{eq:v6}
v_6 &=  \frac{1}{792n(\theta)} \left(\left(3-16 \sqrt{3}\right) \cos \theta +9 \left(25-16 \sqrt{3}\right) \sin \theta \right)\\
\label{eq:tp4}
t'_4 &=  \frac{1}{396n(\theta)} \left(-\left(\sqrt{3}-27\right) \cos \theta -9 \left(\sqrt{3}-5\right) \sin \theta \right) \\
\label{eq:vp6}
v'_6 &=  \frac{1}{792n(\theta)} \left(\left(15+8 \sqrt{3}\right) \cos \theta +9 \left(8 \sqrt{3}-7\right) \sin \theta \right)
\end{align}

\end{document}